\documentclass[twocolumn,showpacs,preprintnumbers,amsmath,amssymb,floatfix,nofootinbib]{revtex4-2}

\usepackage[caption=false]{subfig}  
\usepackage{url}
\usepackage{xcolor}      
\usepackage{graphicx}
\usepackage{subfig}
\usepackage{bm}         
\usepackage{epsfig}
\usepackage{verbatim}
\usepackage{enumerate}
\usepackage{ulem}
\usepackage{float}
\usepackage[columnwise]{lineno}

\begin{document}

\title{Fusion and reactions of $\alpha$+$^8$Be in the Hoyle resonance and associated resonances region}

\author{Teck-Ghee Lee}
\email{tgl0002@auburn.edu; tglee.physics@gmail.com}
\affiliation{Department of Physics, Auburn University, Auburn, Alabama 36849, USA.}

\author{Orhan Bayrak}
\email{bayrak@akdeniz.edu.tr}
\affiliation{Department of Physics, Akdeniz University, Antalya 07058, Turkey.}


\author{Cheuk-Yin Wong} \email{wongc@ornl.gov}
\affiliation{Physics Division, Oak Ridge National
  Laboratory$^1$,
Oak Ridge, Tennessee 37831, USA.}
\footnote
{The research has been supported  in part by
  UT-Battelle, LLC, under contract DE-AC05-00OR22725 with the US
  Department of Energy (DOE). The US government retains the
  publisher, by accepting the article for publication, acknowledges
  that the US government retains a nonexclusive, paid-up, irrevocable,
  worldwide license to publish or reproduce the published form of this
  manuscript, or allow others to do so, for the US government
  purposes. DOE will provide public access to the results of
  federally sponsored research by the DOE Public
  Access Plan (http://energy.gov/downloads/doe-public-access-plan),
  Oak Ridge, Tennessee 37831, USA}

\begin{abstract}
The fusion of  $\alpha$ and $^8$Be to produce a $^{12}$C nucleus is a crucial process in
nucleosynthesis. In the laboratory, this process can only be studied 
theoretically as a $^8$Be target or projectile cannot be prepared experimentally.  
We use the potential scattering theory 
in the coupled-channel formalism
to study such a process in terms
of the collision between the $\alpha$ particle on a deformed
$^8$Be nucleus, both on
resonance and off resonance
in the Hoyle
resonance and associated resonances region. The experimental $^{12}$C energy
levels and widths constrain the nuclear potential to suggest
the need to include a parity-dependent surface potential component that
is more attractive for even-$L$ positive-parity partial waves than for
odd-$L$ negative-parity partial waves.  As a consequence, the radial
dependence of the total potentials for the set of \{0$^+$, 2$^+$, 4$^+$\} resonances
of ${}^{12}$C
exhibit a double-hump behavior, possessing two local energy
minima and a doublet of 
each of the 
 ${}^{12}$C \{0$^+$, 2$^+$, 4$^+$\} 
resonances in the Hoyle and associated resonances region. 
 We examine the approximate agreement of the theoretical results 
 with experiment
and suggest the search for the as-yet unobserved lower-energy
2${}^+_2$ and 4${}_1^+$ resonances to test the
double-hump potential description.
In addition, for practical astrophysical applications, we evaluate and estimate 
the astrophysical $S(E_{\rm c.m.})$-factor for the 
$\alpha$+$^8$Be $\to$ $^{12}$C$(0^{+*})$ $\to$ $^{12}$C$(2_1^+)$ + $\gamma$ 
reaction for $E_{\rm c.m.}$ $<$ 1.0 MeV.
\end{abstract}


\maketitle
\thispagestyle{empty}

\def\centering{\relax}

\section{Introduction}

The fusion between an $\alpha$ particle and a $^8$Be nucleus, leading to the production of the $^{12}$C nucleus,  
is a crucial process in stellar nucleosynthesis, as it is the gateway to 
the production of heavier elements in the Universe. The $^{12}$C  production, 
via the process of Salpeter \cite{Sal52}, and Opik \cite{Opi51}, is only possible 
by a resonance postulated by Fred Hoyle \cite{Hoy54, Hoy53}, now known 
as the {\it Hoyle resonance} ($0_2^+$). This process is referred 
to here as the {\it Salpeter-Hoyle} mechanism \cite{Sal52, Opi51, Hoy54, Hoy53}. 
The Hoyle resonance occurs at the low collision energy of $E_{\rm c.m.}$ = 0.281 MeV, 
which can be reached by the colliding nuclei in a stellar environment with sufficiently high temperatures.   

We study here the Salpeter-Hoyle reaction, $\alpha$ +
$^8$Be $\rightarrow$ $^{12}$C*, using potential scattering theory \cite{Bla52}
in the coupled-channel framework \cite{buc63,Tam65,Tho88,Hag22},
both on resonances and off resonances.
 We can thereby examine 
 the resonance energy profile  due to
the interference of barrier penetration and the propagation of outgoing
scattering waves. This interference will lead to notable modifications
to the Breit-Wigner energy distribution of the Hoyle resonance, to turn it, for
example, into a Breit-Wigner-Fano resonance profile \cite{Fano61,
  Coo65}.  The potential
scattering theory also provides the proper framework to describe the Hoyle
resonance as a pocket resonance in a
potential pocket with a potential barrier \cite{Whe59,Lee21}. 
Such resonances occur commonly 
in atomic collisions \cite{Ovc06}, molecular collisions \cite{Zho20}, 
and, as demonstrated here, also in nucleus-nucleus collisions.  
 They are expected to occur prominently in collisions 
of $\alpha$-conjugate nuclei, ($\alpha$)$^{n_1}$+($\alpha$)$^{n_2}$.   
Furthermore, the resonances obtained from the potential
scattering theory complement those from structure
studies. Finally, such a theoretical study is highly desirable, as an experimental 
measurement of the fusion cross sections for the $\alpha$ particle on 
a $^8$Be target or projectile is
impossible due to the short life time of the $^8$Be$_{\rm gs}$
nucleus. The quantitative study of the $\alpha+^8$Be fusion to $^{12}$C 
in nucleosynthesis must rely on theoretical predictions.

There has been extensive experimental and theoretical computational work devoted to the
understanding of the structure of the Hoyle state and its associated excited states in $^{12}$C \cite{Fre14, Fre18}. 
The adiabatic hyperspherical method with complex scaling for solving the 3$\alpha$-cluster Faddeev equations (HFE), 
developed by Jensen and collaborators \cite{Jen96, AR07, AR07a, AR08}, together with related approaches 
by Descouvemont and collaborators \cite{Des87, Des21, Des12, suno15}, 
has successfully reproduced the theoretical state energies and $\alpha$-decay widths of low-lying 
states in agreement with experimental observations for many $^{12}$C resonances, while microscopic 
$R$-matrix calculations by Neff and Feldmeier \cite{Fel14} have provided further valuable insights.
More recently, {\it ab initio} few-body lattice calculations have shown that the configurations of the 0$_2^+$ Hoyle state 
and the second excited $2_2^+$ state resemble a bent-arm triangle, with two $\alpha$ particles closer together 
and more distant from the third$-$a finding consistent with other studies supporting the 
exotic “molecular” structure of $^{12}$C \cite{Epe12, Vas12, Whe37, Tho13}. 

On a different front, theoretical investigations of the $\alpha+^8$Be radiative  capture (or radiative fusion) 
leading to the compound  $^{12}$C nucleus have also advanced. For example,  Ogata {\it et al.} \cite{Oga09}
performed three-body continuum-discretized coupled-channel (CDCC) calculations that accounted for both resonant 
and nonresonant mechanisms in a consistent manner in the triple-$\alpha$ reaction, showing 
that $\alpha$-$\alpha$ continuum states below the 92.08 keV resonance play a crucial role in 
the nonresonant triple-$\alpha$ process at stellar temperatures below approximately 0.04 GK. 
Subsequently, Suno {\it et al.} \cite{suno15} applied an adiabatic hyperspherical method 
with complex absorbing potential (HCAP) to solve the three-body Schr\"{o}dinger equation, obtaining the cross section for 
the photodisintegration process $^{12}$C$(2^+)$ + $\gamma$ $\to$ $\alpha+\alpha+\alpha$ at astrophysical energies ( $<$ 1.0 MeV)
and the corresponding triple-$\alpha$ reaction rate over the stellar temperature range 0.01$-$1.0 GK. More recently, 
Depastas {\it et al.} \cite{Bon24} sought to evaluate the fusion cross section at comparable low energies
using a statistical 3$\alpha$ model that combines the imaginary time method with a semi‑classical hybrid 
$\alpha$‑clustering and neck model.

We wish to study the Salpeter-Hoyle reaction by employing  the potential scattering theory in a coupled-channel framework \cite{buc63,Tam65,Tho88,Hag22}
that takes into account the deformation of $^8$Be$_{\rm gs}$. The phenomenological optical 
model potential (OMP) parameters are judiciously chosen and constrained by the experimental 
energies and widths of the Hoyle state and its neighboring excited $^{12}$C states. In this work, 
we restrict our focus to the Hoyle resonance and the associated natural-parity 
resonances, $\pi = (-1)^L$, up to $E_x \approx 15$ MeV, where $E_x$ and 
the collision energy $E_{\rm c.m.}$ are related by 
$E_x = E_{\rm c.m.} + 7.367$ MeV. 

It is pertinent to check whether the instability of $^8$Be would significantly affect the 
calculation of the reaction cross section, when we ignore the decay of  $^8$Be$_{\rm gs}$ 
decay  in the treatment of the Salpeter-Hoyle reaction. 
The half-life of the $^8$Be$_{\rm gs}$ is of the order of $10^7$ fm/$c$, which is long 
compared to the timescale of the $\alpha$+$^8$Be fusion \cite{Hoy54}.   It is therefore 
justified to disregard the decay of $^{8}$Be$_{\rm gs}$ in the Salpeter-Hoyle reaction. 
We will also address whether the decay of the $^8$Be(2$^+$) at $E_x =$ 3.03 MeV, 
which has a half-life of the order of $\hbar/$(1.5 MeV) $\approx$ 131 fm/$c$, will affect 
the Salpeter-Hoyle reaction cross sections at $E_{\rm c.m.} < 4.0$ MeV 
in the subsequent discussions in Section VI.

The rest of the paper is organized as follows: Section II outlines our approach to 
determine the potential parameters in the coupled-channel calculations for the $\alpha$+$^{8}$Be collision. 
This section also presents the results of the theoretical reaction cross sections for different partial waves, 
which exhibit complex resonance structures. In Section III, we present a phenomenological 
analysis of the reaction cross sections obtained from the coupled-channel calculations, 
in terms of parameterized resonances and off-resonance reaction cross sections. Section IV is dedicated 
to comparison of the resonance energies and $\alpha$-widths  obtained  in the potential scattering theory 
with those from experiment and  the 3$\alpha$-cluster HFE theory. 
In Section V, we discuss the $^{12}$C $4_1^+$ and $2_2^+$ states at $E_x\approx$ 10 MeV which 
have been predicted by the potential scattering theory but not yet observed or identified. 
We review different and confusing experimental results on the $2^+$ states  at $E_x\approx 10$ MeV 
 and suggest further theoretical and experimental work to clarify the situation.
We also study theoretically the potential landscape  for $L=2$ to provide understanding on the  
$\alpha$-width  $\Gamma_\alpha$ of the $^{12}$C $2^+$ states  at $E_x\approx 10$ MeV.
Section VI presents our evaluation of the astrophysical $S(E_{\rm c.m.})$-factor of 
the $\alpha$+$^8$Be $\rightarrow$ $^{12}$C$(2^+)$ process for energies $E_{\rm c.m.}$ $<$ 1.0 MeV, 
using our $s$-wave $\alpha$+$^8$Be reaction cross sections and the corresponding widths 
of $\gamma$- and $\alpha$-decay for the decay of $^{12}$C excited states in the potential pocket. 
The paper concludes with Section VII, which summarizes our discussions. For brevity of notation, 
unless ambiguity arises, we shall use the symbol $E$ for $E_{\rm c.m.}$, and $^8$Be($0^+$) 
for $^8$Be$_{\rm gs}$ ground state interchangeably throughout the paper.

\section{The optical model potential in the coupled-channel calculation for $\alpha$+$^8$B\lowercase{e} reactions}
The coupled-channel method is a standard tool for analyzing low-energy two-body reactions 
that involve changes in the internal states of the colliding particles. A detailed review of this method is available 
elsewhere \cite{buc63, Tam65, Tho88, Hag22} and is not repeated here. Instead, this section describes how we 
come up with the potential model for the $\alpha$+$^8$Be collisions in scattering calculations 
performed with the FRESCO code \cite{Tho88}.
The total effective optical model potential, experienced by the partial $L$-wave among the coupled $L$ channels 
of the incoming $\alpha$-particle, is given by \cite{Tam65, Tho88, Hag22} 

\begin{eqnarray}
\hspace{-0.1cm} V_{\rm eff}\!(r,\theta)\hspace{0.0cm} &\!=\!&  V_n(r,\theta) \!+\! V_c(r,\theta) \!+\! V^{(L)}_{\pi}(r)+V_L(r),
\label{Vtot}  
\end{eqnarray}
where
\begin{eqnarray}
\hspace{-0.0cm} V_n(r,\theta)  \hspace{0.0cm} &=& -\left(\frac{V_o}{1+e'_v}+\frac{iW_o}{1+e'_w}\right), \label{Vn} 
\\ 
\hspace{-0.0cm} V_c(r,\theta) \hspace{0.0cm}~ &=& \frac{Z_pZ_te^2}{2R_c}\left( \! 3-\frac{r^2}{R^2_c}\right)\Theta(R_c-r) 
\nonumber \\
&&\hspace{-0.60cm}+ \frac{Z_pZ_te^2}{r^2} \Theta(r-R_c) + \hspace{-0.2cm} \sum_{\lambda=2,4,...}\hspace{-0.3cm}\frac{3Z_t \beta_{\lambda}R_c^{\lambda}}{4\pi} \left(\!\frac{\sqrt{4\pi}e^2}{2\lambda+1} \!\right)  
\nonumber \\
&&\hspace{-0.6cm}\times \left(\frac{r^{\lambda}\Theta(R_c-r)}{R_c^{2\lambda+1}} + \frac{\Theta(r-R_c)}{r^{\lambda+1}}\right)Y_{\lambda 0}(\theta'), 
\label{Vc}
 \\
\hspace{0.0cm}V_\pi^{(L)}(r) \hspace{0.0cm} &=& -\frac{(-1)^L\times4 V_{\pi o}\, e_\pi}{(1+e_\pi)^2}, 
\label{Vpi}
\\
\hspace{-0.0cm}V_L\,(r)~&=\!&\! \frac{L(L+1)}{2\mu r^2}, 
\label{VL}
\\
\hspace{-0.3cm}e'_{(v,w)} ~&=&\! \exp[(r\!-\!R_{(v,w)}\{1\!+\!\hspace{-0.2cm}\sum_{\lambda=2,4,...}\hspace{-0.2cm}\beta_\lambda Y_{\lambda 0}(\theta')\})/a_{(v,w)}], 
\nonumber \\ \\
e_\pi~& = & \exp[(r-R_\pi)/a_\pi].
\end{eqnarray}

In Eq.\ (2)-(7), $Z_{\{p,t\}}$ are the nuclear charges for a projectile $p$ and target $t$, 
respectively, $V_\pi^{(L)}(r)$ is the parity- and $L$-dependent surface potential (proportional 
to the first derivative of the Woods-Saxon nuclear potential), $V_{\{n,c\}}$ denote the deformed nuclear ($n$) and, 
to lowest order, deformed Coulomb ($c$) potentials, respectively, where $R_{\{c,v,w,\pi\}}$ = $r_{\{c,v,w,\pi\}}A_t^{1/3}$, $A_t$ 
is the target mass number, $\mu$ is the reduced mass of the collision pair, $Y_{\lambda 0}(\theta')$ 
is the spherical harmonic with the angle $\theta'$ referring to the body-fixed system. Appropriate methods 
to solve the coupled-channel equations were presented in detail in \cite{buc63, Tam65, Tho88, Hag22}. 
The reaction cross section $\sigma_R(E)$ obtained from the coupled-channel calculations can 
be decomposed into the sum of the partial reaction cross section $\sigma_R(E,L)$ in the standard way.

\begin{figure}[htp]
\includegraphics[width=1.00\linewidth,height=1.25\linewidth]{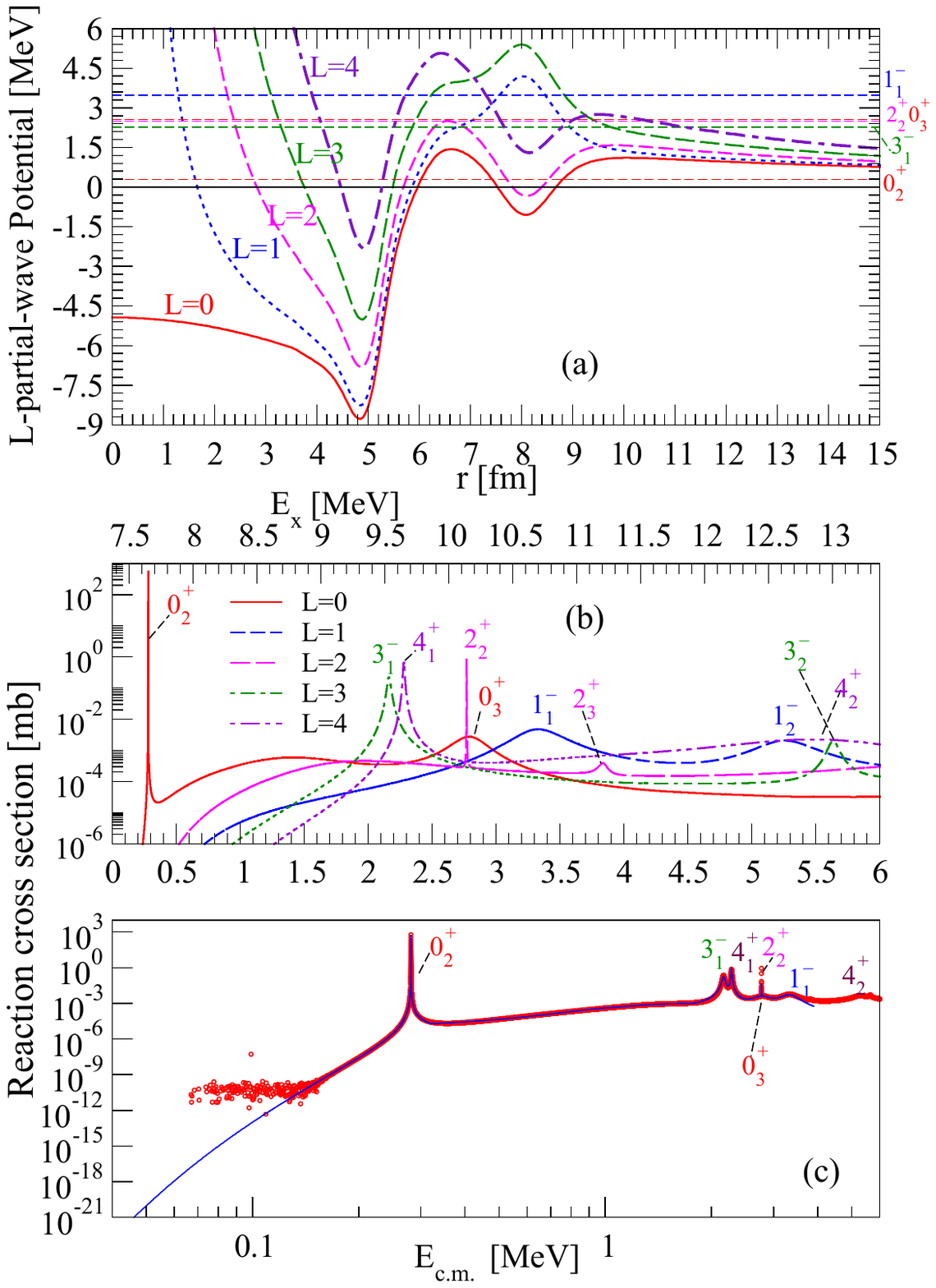} 
\vspace{-0.8cm}
\caption{(a) The real parts of the $\alpha$+$^{8}$Be interaction potentials for different $L$ orbital angular momenta, 
as calculated by the FRESCO code. The horizontal dashed lines mark the corresponding energy levels $E$ of 
the experimental $L^{\pi}$ resonance states of $^{12}$C \cite{Kel17}. (b) $L$-partial-wave reaction cross section $\sigma_R(E_{\rm c.m.},L)$ 
 obtained in the coupled-channel calculations using the FRESCO code,
 with the channel couplings between the $^8$Be(0$^+$) and $^8$Be(2$^+$) states.
(c) The open circle, shown on a log-log scale, is the total $\sigma_R(E_{\rm c.m.})$, summed over all $L$ and the solid curve is the  phenomenological reaction cross section, 
evaluated using Eqs.\ (\ref{BW2}) to (\ref{eq21}) in the analysis of Section III. The coupled-channel calculation 
reaches its lowest energy limit at $E_{\rm c.m.}$ $\approx$ 0.15 MeV.}
\label{fig1}   
\end{figure} 

To search for the potential depth, $V_o$ in Eq.~(\ref{Vn}), we first perform numerous $s$-wave scattering calculations 
with a hypothetical spherical $^8$Be nucleus, at an energy fixed at the Hoyle energy of $E$ = 0.28 MeV. After extensive experimentation, 
we obtained a sequence of central potential well depths $V_o$ = 4.4, 11, 22, 38 and 59 MeV.  It is necessary to choose a 
starting potential depth $V_o$ from this set based on sound physical principles for our problem at hand.

In the study of the nuclear structure of $^{12}$C, Buck {\it et al.} \cite{Buc13} used a deep and local two-body potential 
to model the interaction between $\alpha$ and $^8$Be. They chose a potential depth of $V_o$ = 144 MeV (see Eq.~(2) in Ref.\cite{Buc13}),
combined with the Coulomb potential within the cluster-core Hamiltonian. By solving the Schrödinger equation, 
they obtained the excitation energy levels, widths, and charge radii for the first few states of the Hoyle band in $^{12}$C. 
However, this deep potential of 144 MeV may introduce an artifact: besides the lowest two Pauli-blocked and hard-$\alpha$-core-blocked  
states occupied by the two $\alpha$ particles of the $^8$Be nucleus in the potential, there are many vacant single $\alpha$-particle
bound states that can lead to spurious high-energy $\gamma$ transitions when an $\alpha$ particle forms a compound system 
with $^8$Be in the potential pocket, which can decay down to vacant single $\alpha$-particle bound states by electromagnetic radiation.

\begin{center} 
\begin{table}[ht]
\caption{Parameters of the optical model potential for $\alpha$+$^8$Be system. The deformation parameter $\beta_2$ = 0.6 \cite{Abe94} is used for both nuclear and Coulomb terms in Eq.\ (1). The $V^{\rm ev}_o \equiv V^{(L=0,2,4,...)}_o$, $V^{\rm od}_o \equiv V^{(L=1,3,5,...)}_o$ and $V_{\pi o}$ are in the units of MeV; the $W_o$ is in eV; the radii $r_{\{c,v,w,\pi\}}$ and diffuseness widths $a_{(v,w, \pi)}$ are in fm.}
\begin{tabular}{ p{0.8cm}  p{0.7cm}  p{0.8cm}  p{0.7cm} p{0.6cm}  p{0.6cm}  p{0.6cm}  p{0.6cm}  p{0.6cm} p{0.6cm} 
p{0.58cm}  } 
\hline 
 $V^{\rm ev}_o $ & $V^{\rm od}_o$ & $r_v$ &  $a_v$ & $W_o$ & $r_w$ & $a_w$ &  $V_{\pi o}$ &  $r_\pi$  & $a_\pi$  & $r_c$ \\ 
\hline 
9.698  & 9.86 & 1.469 &  0.18 &  1.5 & 1.47  & 0.4  &  2.5  & 2.26 & 0.37 & 1.01 \\
\hline 
\end{tabular}
\label{tab0} 
\end{table}
\end{center}

To address the problem of finding the proper well depth as a suitable starting point, 
we consider the optical model potential for the $\alpha$+$^8$Be collision to be 
the extension of the single-$\alpha$-particle bound state potential for bound states in the $^8$Be nucleus, 
to a single-$\alpha$-particle {\it resonance-state} potential that accounts for {\it resonance states} 
in the continuum. This extension is achieved by introducing an additional degree of freedom via imaginary potential, $W(r)$.
We envisage that in such an $\alpha$+$^8$Be system leading to a compound $^{12}$C nucleus, the 
lowest $\alpha$-single-particle $1s$ and $1p$ bound states will be occupied by two hardcore $\alpha$ particles, and the incident $\alpha$ particle can settle on the next $2s$ and $1d$ states at low collision energies as resonances, with the $2s$ state prepared to become the Hoyle resonance.  A potential well depth of about 11 MeV will have the lowest $1s$ and $1p$ states as bound states and the $2s$ and $1d$ as resonance states just emerging into the continuum.
Therefore,
we selected 11 MeV as our starting point of our search of potential parameters,
 from the range of potential depths previously identified. 
This choice of $V_o \approx$ 11 MeV 
agrees approximately with the well-depth shown for an attractive mean-field potential for an $\alpha$ particle 
in the $^{12}$C nucleus in a 3-$\alpha$ cluster HFE problem, as described in Ref.\cite{AR07, AR07a, Des21, Fel14}. 
It is noteworthy that, after extensive fine-tuning, we settled on a radius of $r_v = 1.469$ fm, 
which is slightly smaller than the $r_o = 1.6$ fm suggested in Ref. \cite{Fre18}.

The coupled-channel scattering calculation considers the $^8$Be  target as a prolate nucleus.
Through iterative adjustments, we established a deformation parameter $\beta_2$ = 0.6 \cite{Abe94} 
and other OMP parameters. First, we kept the potential depth at $V_o$ = 11 MeV to refine the geometric 
parameters in the coupled-channel calculations. This process continued until we achieved a set of satisfactory 
resonances for energies $E$ below 4 MeV. However, 
to get our results agree reasonably well 
with the experimental properties of the Hoyle and other $^{12}$C resonances,
we need to introduce a parity-dependent surface potential component $V_\pi^{(L)}(r)$ in Eq.\ (1) 
that is attractive for the  even-$L$ positive-parity states
and repulsive for the odd-$L$ negative-parity states.

The need for a phenomenological parity-dependent surface potential component in our two-body interaction potential 
may be attributed to the effects of the {\it Bose-Einstein exchange} in boson system 
in spatially extended configurations. The ground state of such a boson system is symmetric 
with respect to the exchange of identical bosons, leading to lowered surface potential energy for 
even-parity interactions within the $\alpha$ and $^8$Be potential. Note that the concept 
of a parity-dependent surface potential component for the scattering of $\alpha$ particles has been previously 
introduced and demonstrated by several authors \cite{Kon75, Kan63, Tan72} to analyze and
explain the backward-angle anomaly observed in the differential cross section 
for $\alpha$ scattering on $\alpha$-conjugate nuclei.

This concept of the parity-dependence of the surface potential component also appears to be analogous to the parity-dependence of the  strength of the phenomenological 
attractive effective three-body interaction
$S$ in the 3$\alpha$-cluster HFE theory which  treats $^{12}$C as a three-$\alpha$ cluster
system in Ref.\ \cite{AR07}. There, $S$ parametrizes the depth of their 3-body interaction, expressed as 
$V_{3b}(\rho) = S \exp(-\rho^2/b^2)$, where $\rho$ is the hyperradius and $b$ was chosen to be approximately 6.0 fm such that when $\rho$ = $b$ all three $\alpha$ particles will be touching. Table 2 of Ref.\ \cite{AR07} showed that the three-body interaction $S$ is more attractive  for even-parity states than for odd-parity states, with $S$ taking the form $S$ = $S_0 + (-1)^LS_1$, 
where $S_0 \approx -11$ MeV and $S_1 \approx -9$ MeV. 

The calculation of the resonance energies and widths requires precision, and  necessitates  judicious and careful
adjustments of the potential parameters. Systematic fine-tuning is essential for a reasonable 
description of the low-lying Hoyle resonance and other natural parity  $L^pi$ resonances. 
Fig.\ref{fig1}(a) shows  the real part of the total effective 
potential for different $L$ orbital angular momenta and parity $\pi=(-1)^L$, obtained using the FRESCO code with 
the optical model potential parameters for $\alpha$+$^8$Be system listed in Table \ref{tab0}. These 
curves provide insights into the behavior of the system. 

In Figure \ref{fig1}(a) we show the experimental energy level of an $L^{\pi}$ resonance plotted with its corresponding  total effective potential to find out whether the $L^\pi$-resonance 
qualifies as a sharp pocket resonance (energy level lying below the potential barrier)  \cite{Lee21} or a broad Ramsauer-type resonance 
(energy level lying considerably above the barrier peak) \cite{Ram21, Pet62, Mar68}. To locate these resonances accurately, as depicted in the plot of the partial 
reaction cross sections versus energy plot Fig.\ref{fig1}(b), we consider the scattering boundary conditions at a matching radius of 50 fm. 
A 22,000 points energy grid with varying step sizes was employed to resolve these remarkably narrow resonances. 
By fitting the reaction cross-section data with either a Breit-Wigner or a Breit-Wigner-Fano model, we extract the theoretical 
resonance energies and widths, which will be discussed in details in the subsequent sections. 

Of all the resonances presented in Fig. \ref{fig1}(b), 
the Hoyle resonance at the excitation energy of $E_x$ = 7.648 MeV relative to the ground state of $^{12}$C, and the broad 0$_3^+$ resonance 
at $E_x$ = 10.16 MeV, serve as critical benchmarks in our coupled-channel optical model analysis. 
Note that extensive numerical tests on the present potential
scattering model have indicated that the resonant width,
$\Gamma_\alpha$ and the reaction cross section are highly sensitive
to the shape of the potential curves and $W_o$.  The parameter $W_o$
 represents effects of absorption arising
from channels that have not been explicitly included in the
calculations.  We calibrate $W_o$ by requiring that reaction cross
sections at energies far from the resonance match those as predicted
by the Gamow-Sommerfeld barrier penetrability of Eq.\ (\ref{Pcont}) 
as shown later in Fig. \ref{fig3}(b) below, with the value of $W_o=1.5 $ eV.

\section{Phenomenological description of coupled-channel $\alpha$+$^8$B\lowercase{e} reaction cross sections}
This section describes the results of the potential scattering theory in coupled-channel 
calculations and examines the underlying physics. 
In Figs.\ \ref{fig1}(b) and (c),  we show  the reaction cross section $\sigma_R$ as a function 
of the center-of-mass energy $E_{\rm c.m.}$ of $\alpha$+$^{8}$Be, or equivalently, 
the corresponding excitation energy $E_x$ of $^{12}$C$^*$. These results
exhibit a complex structure  which  requires a comprehensive description.

From the complex structures of the $\sigma_R$, 
we can carry out first a phenomenological analysis
to extract important information on $^{12}$C 
resonances, both at resonance and off-resonance. 
The extracted resonance energies and widths of these resonances can be compared directly with the 
corresponding $^{12}$C experimental data and calculations from 3$\alpha$-cluster HFE theory \cite{AR08}.  

Secondly, we wish to generalize our previous analysis of ``off-resonance'' reaction cross section in \cite{Won73} 
on barrier penetration in the absence of resonances to the present case which involves barrier penetration in the presence of resonances.
This will allow us to study  
 interference between 
the resonance and the underlying barrier penetration, similar to the interference between potential 
pocket resonances and barrier penetration in a {\it double-hump} barrier \cite{Won69,Bra72}.

Thirdly,  we also wish to extract 
information on the probability of potential barrier penetration at the energy $E_{\rm c.m.}$  for the $L$ partial wave, $P(E,L)$,  needed for the calculation of the $\alpha$-decay width $\Gamma_\alpha$, 
as given later in Eq.\ (\ref{eq7}).   The $\alpha$-particle penetrability for going through a realistic 
potential barrier would otherwise 
be challenging to acquire. 

Finally, for a given set of the spatial integration region and the integration grid,
the coupled-channel approach can reach its lowest energy numerical limit, 
below which the calculated results become impractical due to numerical errors 
coming from the finite sizes in both the spatial integration 
region and the spatial  integration grids, as shown in Fig.\ 1(c) for $E \lesssim 0.15 $ MeV. 
However, the energies below this lowest energy limit fall within the range of significant interest 
in nucleosynthesis studies, particularly in assessing the fusion cross section of compound nucleus $^{12}$C 
in $\alpha$+$^8$Be collisions. For this reason, it is necessary to express the numerical 
cross sections in analytical phenomenological forms, extending to the lowest energy possible 
that are inaccessible through a coupled-channel or other theoretical methods, for both theoretical 
understanding and practical applications.
\begin{figure}[h]
\includegraphics[scale=0.35]{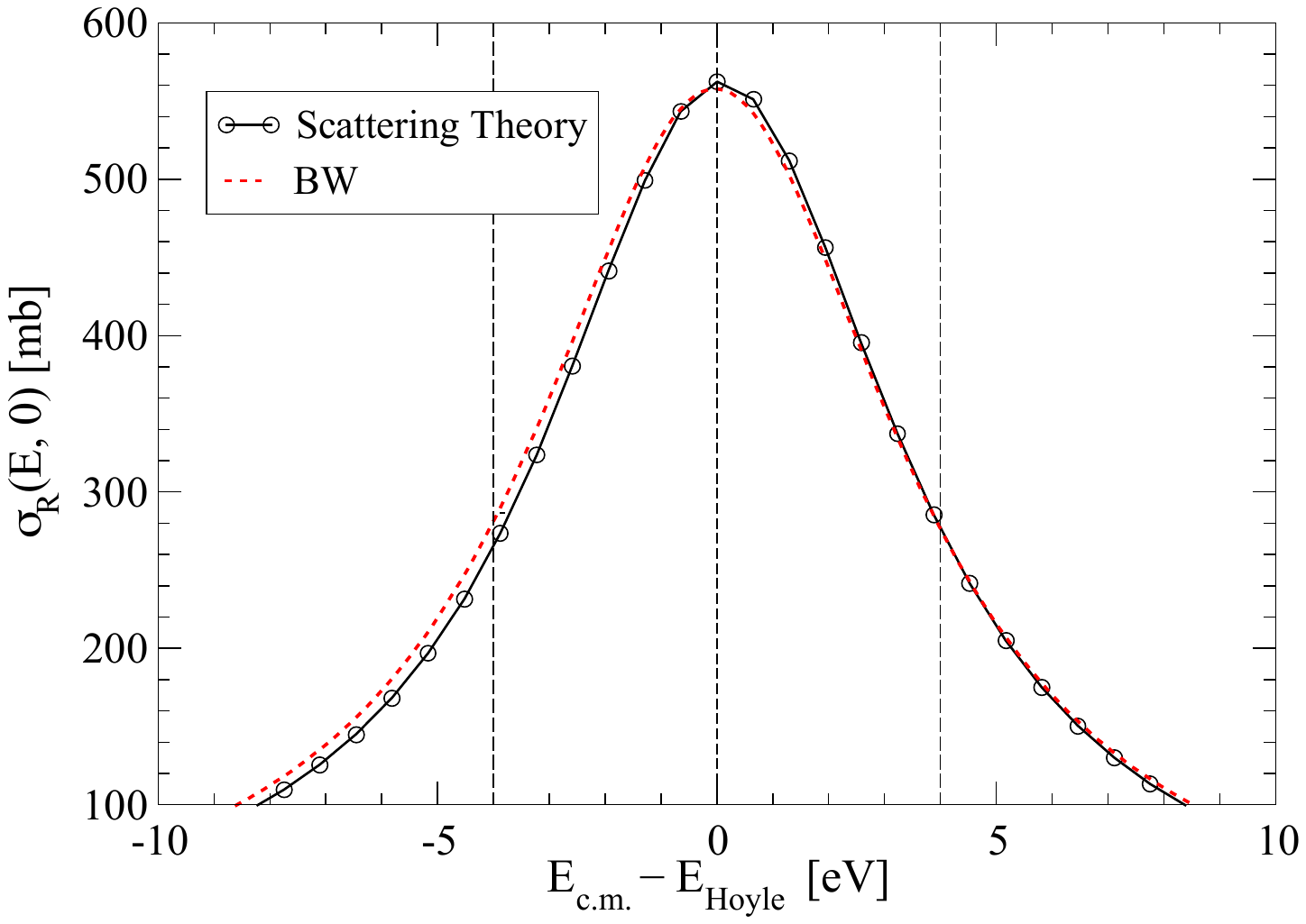}
\vspace{-0.8cm}
\caption{The solid curve with circles is the
$s$-wave reaction cross section at the Hoyle resonance  obtained from  
the potential scattering theory in coupled-channel calculations.
The dotted curve is the parametrization of the $s$-wave reaction cross section as
a Breit-Wigner resonance,  as $\sigma^{(L=0)}_R(E)= (\pi/k^2)(2/\pi)P_{\rm res}(E_\zeta,L=0)/[(E-E_\zeta)^2/(\Gamma_\zeta/2)^2+1]$, where $E_\zeta = E_{\rm Hoyle}$ = 0.2809338737 MeV, $\Gamma_\zeta=8.0$ eV and $P_{\rm res}(E_\zeta, L=0)= 1.0$.}
\label{fig2}
\end{figure}

The reaction cross section $\sigma_R$ depicted in Fig. \ref{fig1}(b) for different $L$ partial-waves demonstrates 
resonance structures against a background of a relatively smooth variation as a function of $E_{\rm c.m.}$. 
The energy region, denoted as $\cal R$,
can be conveniently and conceptually partitioned into two distinct regions: the resonance regions, called $\cal R_{\rm res}$, 
which surround the resonances, and the off-resonance, continuum regions,  called $\cal R_{\rm contin}$, 
located between the resonances. These two regions are governed by different physical principles, 
allowing for a judicious division of the reaction cross sections.

Within the resonance energy regions,  ${\cal R}_{\rm res}$, we envisage that the distinctively sharp resonances 
could manifest in the reaction cross sections $\sigma_R$. These are linked to pocket resonances.
Additionally, there may be resonances characterized by large widths, known as  above-the-barrier Ramsauer-type resonances. 
These arise due to the interference between waves traversing through the nucleus and those around the nucleus \cite{Ram21, Pet62, Mar68}.   

In the off-resonance continuum region ${\cal R}_{\rm contin}$, the states of the
$^{12}$C compound nucleus lies in the continuum. The  $\sigma_R$ between resonances in this ${\cal R}_{\rm contin}$ region  
depends not just on the ability to penetrate the potential barrier but also by how the wave function is matched 
inside the potential pocket at the continuum state. Consequently, the reaction cross section depends sensitively also on  
the shape and the barrier height of the potential.   The resonance regions $\cal R_{\rm res}$ disperse themselves among  the off-resonance regions of $\cal R_{\rm contin}$ of the continuum.

The reaction cross section at an energy $E(=E_{\rm c.m.})$ in a nucleus-nucleus collision 
can be expressed, in general, as Ref. \cite{Won73},
\begin{eqnarray}
\sigma_R(E)  
= \sum _{L=0} \sigma_R^{(L)} (E)
=  \sum _{L=0}\frac{(2L+1)\pi}{k^2(E)}  P(E,L), 
\end{eqnarray}
where $P(E,L)$ represents the probability  for the incident particle 
with partial wave $L$ to commence from the external entrance point beyond the barrier,
penetrate the potential barrier, and reach the interior potential pocket region of the target nucleus. 
We can decompose the contributions from states in the entire energy region as an integral over 
individual state specified by the Dirac $\delta$-function at different energies $E$ as
\begin{eqnarray}
\sigma_R(E)  
= \sum _{L=0} \int_{\cal R}   dE ' \delta (E -E')  \frac{(2L+1)\pi}{k^2(E')}  P(E',L).
\end{eqnarray}

As $\cal R $=$ \cal R_{\rm res}$+$ \cal R_{\rm contin}$,  we have 
\begin{eqnarray}
\sigma_R(E) 
\!\!= \!\!\sum _{L=0} \biggl \{ 
\int_{{\cal R}_{\rm res}} \!  dE ' 
 \delta (E -E') 
\frac{(2L+1)\pi}{k^2(E')}  P_{\rm res}(E',L)
\nonumber\\
+ \int_{{\cal R}_{\rm contin}}  \hspace*{-0.6cm} dE '
 \delta (E -E') 
\frac{(2L+1)\pi}{k^2(E')}  P_{\rm contin}(E',L)\biggr \},~~
\label{eq12}
\end{eqnarray}
where $P_{\rm res}(E,L)$ and $P_{\rm contin}(E,L)$ are introduced as the phenomenological penetrabilities 
in the resonance region and off-resonance continuum region, respectively. The penetrability  $P_{\rm res}(E,L)$ 
for a quasi-bound state is expected to exhibit a resonance structure and its energy position depends predominantly 
on the pocket potential satisfying the 
Bohr-Sommerfeld quantization rule. The width of this resonance is determined by the 
potential barrier penetration or the tunneling process from the potential pocket to the outside region. 
The penetrability $P_{\rm contin}(E, L)$ for a continuum state, on the other hand, is expected to exhibit 
a smooth energy-behavior as it depends on the penetration over the potential barrier from the pocket region 
to the outside region. Nevertheless, in the region near a pocket resonance, there may be interference between 
the barrier penetration and  the presence of the pocket resonances. This is because the barrier 
penetrability $P_{\rm contin}(E,L)$ is also dependent on the matching of the wave function at the 
boundary of the potential pocket.

In the off-resonance continuum region,    
the integral $\int_{{\cal R}_{\rm contin}}\!\!dE' ...$ can be simply integrated out to yield 
the partial reaction cross section for the $L$-partial wave  
for low-energy $\alpha$+$^8$Be(0$^+$)$\rightarrow$$^{12}$C$^*$ reactions, 
\begin{eqnarray}
&&\sigma_R^{(\rm contin)}(\alpha+ {}^8{\rm Be}(0^+)\to{}^{12}{\rm C}^*,E, L)
\nonumber\\
&&= \int_{{\cal R}_{\rm contin}} dE'  \delta(E - E') \frac{(2L+1)\pi}{k^2(E')} P_{\rm contin}(E',L) 
\nonumber\\
&&= \frac{(2L+1)\pi}{k^2(E)} P_{\rm contin}(E,L) .
\label{contin}
\end{eqnarray}
As an example of $P_{\rm contin}(E, L)$ is the Hill-Wheeler's barrier-penetration probability (or penetrability)
for an inverted parabola, as given in Eqs. (1) and (2) of Ref. \cite{Won73}. For our case, based on the results of the 
present potential scattering theory, the penetrability will depend not only on the potential 
barrier beyond the potential pocket at large $r$, but also on the matching of the wave function inside 
the potential pocket. 

Within the resonance regions  ${\cal R}_{\rm res}$, the region ${\cal R}_\zeta$ 
for the $\zeta$-th resonance is characterized by a resonance energy $ E_\zeta$ and an width $\Gamma_\zeta$ for the $L$ partial wave, and 
the delta function distribution  in Eq.\ (\ref{eq12}) should be modified to
adopt a Breit-Wigner distribution  $\Delta_{\rm BW}(E-E')$ with a width $\Gamma_\zeta$,  
\begin{eqnarray}
\delta (E - E' )\bigg |_{E'\to E_\zeta} \!\!\! &\to&\Delta_{\rm BW}(E - E_\zeta ), \nonumber
\end{eqnarray}
where
\begin{eqnarray}
\hspace{0.4cm}\Delta_{\rm BW}(E - E_\zeta )\!\!&=&\!\! \frac{2}{\pi \Gamma_\zeta} \frac{1}{[(E-E_\zeta)/(\Gamma_\zeta/2)]^2 + 1}.
\end{eqnarray}

In the present treatment of the $\alpha$ reaction process, the only channel considered is the interaction between the $\alpha$ particle and the $^8$Be nucleus 
modeled by a deformed nuclear potential. Therefore the width $\Gamma_\zeta$ extracted here is implicitly an $\alpha$-width, $\Gamma_{\alpha \zeta}$ for the $L^{\pi}$ state at $E_\zeta$ of partial wave $L_\zeta$ that is predicted by the potential scattering theory.   
The  Breit-Wigner distribution  $\Delta_{\rm BW}(E-E_\zeta)$ is normalized by 
\begin{eqnarray}
\hspace*{0.2cm}\int \!\!dE \Delta_{\rm BW} (E - E_\zeta )&=&\frac{2}{\pi \Gamma_\zeta} \! \!\int \!\!\!
\frac{dE}{[(E-\!E_\zeta)/(\Gamma_\zeta/2)]^2 \!+ \!1}
\nonumber\\
&=& 1.
\end{eqnarray}

For very narrow resonance such as the Hoyle resonance, we can choose ${\cal R}_{\rm res}$ to be only in the bin at $E_{\rm res}$ with the width $\Gamma_\zeta$, with the resonance amplitude given by Eq. (13), and  $\Theta({\cal R}_\zeta)=\Theta (E_\zeta+\Gamma_\zeta/2)- \Theta (E_\zeta-\Gamma_\zeta/2)$. 
Thus in the neighborhood of the resonance $E_\zeta$ for a partial wave $L$, the reaction cross section is therefore
\begin{eqnarray}
&&\sigma_R^{({\rm res})}(\alpha+ {}^8{\rm Be}(0^+))\to{}^{12}{\rm C}^*(E, L)  
\nonumber\\
&&\hspace*{-0.1cm}= \int_{{\cal R}_\zeta} dE'  \delta(E - E') \frac{(2L+1)\pi}{k^2(E')} P_{\rm res}(E',L) 
\nonumber\\
&&\hspace*{-0.1cm}= \Gamma_\zeta   \Theta ({\cal R}_\zeta) 
\left [ \Delta_{\rm BW}(E - E') \frac{(2L+1)\pi}{k^2(E')} P_{\rm res}(E',L) \right ]_{E' = E_\zeta}  \nonumber\\
&&\hspace*{-0.1cm}= \frac{(2L+1)\pi}{k^2(E)}  \left(\frac{2}{\pi}\right) 
\frac{P_{\rm res} (E_\zeta,L)  \Theta ({\cal R}_\zeta)}{[(E-E_\zeta)/(\Gamma_\zeta/2)]^2 + 1}.
\label{eq17}
\end{eqnarray}

Figure \ref{fig2} shows how well Eq.(\ref{eq17}) with $P_{\rm res}(E, L$=$0)$ = 1 
and the normalized Breit-Wigner distribution can describe the $s$-wave 
reaction cross section in the neighborhood of the Hoyle resonance obtained from the 
potential scattering theory in the coupled-channel calculations, locally on the scale of eV energies.  
The good description suggests that the $s$-wave resonance is without substantial absorption.  
An analogous example of the penetrability in pocket resonances can be found 
in Fig.\ 1 of Ref.\cite{Won69}, where the potential between a double-hump barrier leads to pocket 
and resonance structures in the large penetrability at the resonance energies. In addition, 
Fig.\ \ref{fig2} also shows that, at the Hoyle resonance, the $\alpha$ width $\Gamma_{\alpha} \approx \Gamma_\zeta$ = 
8.0 eV is close to the most recent measured value of 8.5 eV \cite{Fre14}. Of course, further fine-tuning 
the two-body interaction potential could further improve our value to  match the experimental width better. 

For resonance with a broad width we can use Eq.\ (13) to represent the resonance 
for all energies and allow $\Theta(\cal R_{\rm res})$ to extend over the entire energy 
region without restriction. This is because the amplitude in Eq.\ (13) decreases rapidly 
beyond a few units of the resonance width $\Gamma_\zeta$, so such an extension will not incur large errors.  

When all the resonances in the ${\cal R}_{\rm res}$ resonance region are included, 
the reaction cross section can be expressed as 
\begin{eqnarray}
&&\sigma_R^{({\rm res})}(\alpha+ {}^8{\rm Be}(0^+))\to{}^{12}{\rm C}^*(E)  
\nonumber\\
&&= \sum_{\zeta=1}^{\zeta_{\rm max}} \frac{(2L_\zeta +1)\pi}{k^2(E)}  \left(\frac{2}{\pi}\right) 
\frac{P_{\rm res} (E_\zeta,L_\zeta )  \Theta ({\cal R}_\zeta) }{[(E-E_\zeta)/(\Gamma_\zeta/2)]^2 + 1}.
\label{BW2}
\end{eqnarray}
The resonances energies $E_\zeta$, resonance width $\Gamma_\zeta$, and 
partial-wave $L$, and the penetrability $P_{{\rm res}} (E_\zeta,L)$ extracted from the potential scattering theory obtained
 $\sigma_R$ are tabulated in Table \ref{tab1}. 

\begin{center} 
\begin{table}[ht]
\caption{The parameters $P_{\rm res} (E_\zeta,L)$, $\Gamma_\zeta$, and $E_\zeta$ in Eq.\ (\ref{BW2}) obtained  from the phenomenological fits to the  numerical results of the
reaction cross sections of potential scattering coupled-channel calculations. 
For the purpose of display, the calculated $E_{\rm Hoyle}$ = 0.2809338737 MeV is truncated to $E_1$ = 0.281 MeV.}
\begin{tabular}{p{0.75cm}  p{1.5cm}  p{2.5cm}  p{1.5cm}  p{1.5cm} } 
\hline\hline 
$\zeta$  & $L^{\pi}$ & $P_{\rm res} (E_\zeta,L)$ & $E_\zeta$/MeV &  $\Gamma_\zeta$/keV \\ [0.5ex] 
\hline 
1 & 0$_2^+$ & 1.0 & 0.281 &   0.008  \\
2 & 3$_1^-$ & 3.5$\times10^{-4}$ & 2.165    &   38 \\
3 & 2$_2^+$ & 4.868 $\times10^{-1} $                      & 2.770     &   0.026  \\
4 & 2$_3^+$ & 1.885$\times10^{-6}$    & 3.837     &   130  \\
5 & 0$_3^+$ & 5.655$\times10^{-4}$ & 2.780    &    300 \\
6 & 1$_1^-$ & 3.56$\times10^{-5}$  & 3.330    &   409  \\
7 & 4$_1^+$& 1.257$\times10^{-3}$    & 2.279    &   12.4 \\
8 & 1$_2^-$& 2.25$\times10^{-5}$    & 5.270    &   610 \\
9 & 4$_2^+$& 8.3776$\times10^{-6}$    & 5.473    &   1650 \\
10 & 3$_2^+$& 1.0$\times10^{-5}$    & 5.637    &   95 \\
\hline 
\end{tabular}
\label{tab1} 
\end{table}
\end{center}

\begin{figure}[h]
\centering
\includegraphics[scale=0.4]{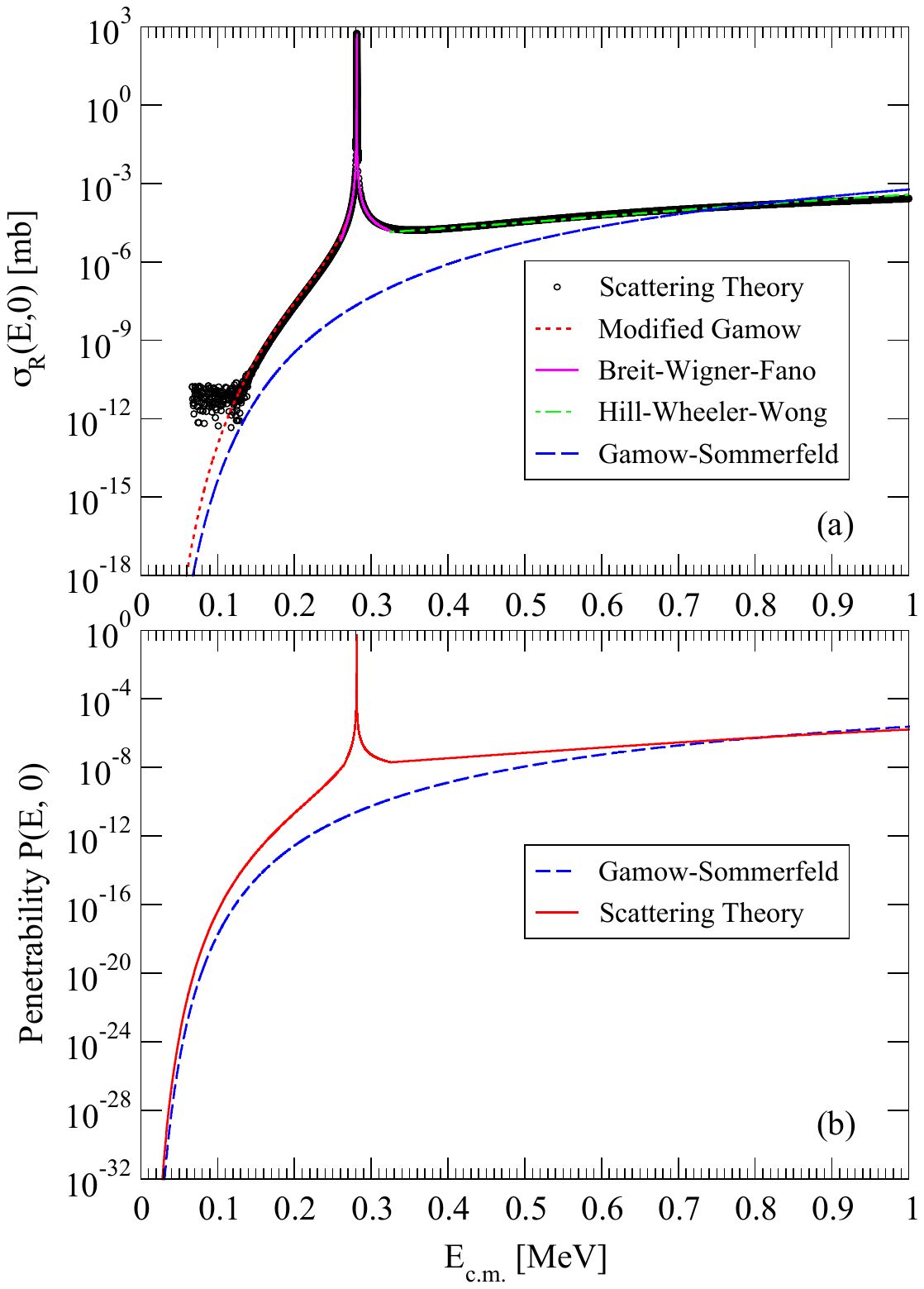}
\vspace{-0.8cm}
\caption{(a) $s$-wave reaction cross sections as a function of energy  for $E_{\rm c.m.}< 1$ MeV
on a semi-logarithmic scale. 
The circle points give  the numerical results from the potential scattering theory in coupled-channel calculations, the curves are various fits and extrapolation  to the circular points. 
(b) The solid curve represents the phenomenological penetrability $P(E, L=0)$, 
extracted from the scattering theory coupled-channel calculations,
as a function of energy on a semi-log scale.  The dashed curve is the Gamow-Sommerfeld penetrability given by Eq. (\ref{Pcont}).}
\label{fig3}
\end{figure}

There is however an important amendment to the Breit-Wigner distribution in the 
presence of barrier penetration. As shown in Fig.\ 2, 
the Breit-Wigner distribution is a good representation to characterize 
a resonance  
within a localized energy scale of the order of the width of the resonance. However, 
when the resonance is accompanied by barrier penetration, as in the case of a pocket resonance,
important effects emerge  arising from 
the interference between the resonance and the 
the underlying continuum barrier penetration, over an energy-scale large compared to the width of the resonance.
As energy changes and the energy level relative to the top of the potential barrier changes,
the penetrability over the barrier also changes.
Furthermore, it  is necessary to have wave functions matching at the boundary of the potential well.   The presence of the quasi-bound state 
enhances penetration into the potential well 
 at the resonance energy and  near the resonance energy over a larger energy scale, as shown by comparing the results of the scattering 
 theory penetrability with the penetrability from Eq.(\ref{Pcont}) in Fig.\ 3(b).

Specifically for the  $s$-wave,
the scattering theory coupled-channel calculations gives locally a Breit-Wigner distribution that is symmetric  with respect to $E-E_{\rm Hoyle}$ with a narrow 8-eV width, 
as shown in Fig.\ \ref{fig2}. However, over a large energy scale of order MeV, 
the 
result of the potential scattering theory coupled-channel calculations 
shows a significantly broadened distribution with asymmetry and a change in shape over a large MeV energy scale, exhibiting higher 
penetrability in the higher collision energy region relative to the lower collision energy region.  The penetrability is considerably enhanced over the Gamow distribution, 
as illustrated in Fig.\ 3. Therefore, within the energy range of 0.26 $\le$ $E_{\rm c.m.}$ $\le$ 0.32 MeV, the $P(E, L)$ 
probability of the $\sigma_R(E,0)= \pi P(E,0)/{k^2}$ reaction cross section has to be modified 
to the Breit-Wigner-Fano resonance profile \cite{Fano61, Coo65} by introducing 
an additional energy-scaled $q\Gamma_\zeta/2$ with dimensionless $q$ to 
represent the $s$-wave reaction cross section near the Hoyle resonance:
\begin{eqnarray}
&&\sigma_R^{({\rm res})}(\alpha+ {}^8{\rm Be}(0^+)\to{}^{12}{\rm C}^*,E, L=0)  
\nonumber\\
&&\hspace*{-0.3cm}\approx\! \frac{\pi}{k^2(E)}  \!\left(\!\frac{2}{\pi} \!\right)\!
\frac{ P_{\rm res} (E_\zeta,L=0)\Theta ({\cal R}_\zeta)}{[(E-E_\zeta)/(\Gamma_\zeta/2)]^2 + 1}
\!\!
\left[\frac{E-E_\zeta}{q\Gamma_\zeta/2} +1\right ]^2 \!\!\!\!,~~
\label{BWF}
\end{eqnarray}
where $E_{\rm Hoyle}=$ 0.2809338737 MeV, $\Gamma_{\rm Hoyle}=8$ eV, 
$q^2 = 1.5 \times 10^8$ and $P_{\rm res} (E_{\rm Hoyle},L$=$0)$ $\approx$ 1. 

The finding sheds light on the significant non-perturbative effect
of barrier penetration  on the energy profile of the resonance over a large energy scale.  Consequently, it becomes necessary 
to introduce an additional energy-scale parameter $q\Gamma_{\rm Hoyle}$= 98  keV  associated with the potential barrier  penetration
in the modified Breit-Wigner-Fano distribution of Eq.(\ref{BWF}), as shown in Fig.\ 3(a).
This inclusion is crucial to represent the
interference between the resonance and the continuum states under the barrier, 
as substantiated by the results of the potential scattering theory, 
particularly within the energy interval of 0.26 $\le$ $E_{\rm c.m.}$ $\le$ 0.32 MeV.  While this example is for the $s$ partial wave, the other resonance with different $L$ values will be similarly modified.

It is noteworthy that the Fano modification of the Breit–Wigner energy profile takes place over an energy scale much larger than the resonance width.
Such a modification is essential for the $s$-wave because other partial waves contributions are negligible. 
It may also be important if the contributions of that partial wave can be isolated over a large energy scale compared to its width.  For other situations when many 
contributions are present, the modification in the distant-energy region 
will be small.
 For example, in the reaction cross section at energies beyond the very low energy region, many partial waves contribute, and the Fano modification becomes less important.  Therefore, it is reasonable to use the Breit-Wigner energy profile 
for all but the $s$-wave ($L=0$) resonances with $\zeta$ = 2, 3, 4,... as presented in Table \ref{tab1}. 
Furthermore, the cross sections 
in the resonance region decrease so rapidly relative to the continuum background that the total reaction can be 
approximated by adding the resonance contributions onto the background distribution, without noticeable differences.
Consequently, for regions beyond the Hoyle resonance, we apply the Breit-Wigner fitting to the remaining resonances 
ranging from 1.77 to 4 MeV.  

We explain how the cross section in the off-resonance
continuum region ${\cal R}_{\rm contin}$ can be described by analytical expressions. With the Hoyle resonance located at $E=0.281$ MeV and the other resonances starting to overlap at  $E \ge 2$ MeV, the continuum  region can be considered to consist of the regions $E\le 0.26$ MeV and $0.32\le E \le1.77$ MeV. 

For  the low energy region, the classic Gamow reaction cross section is \cite{Ian} 
\begin{eqnarray}
\sigma_R^{({\rm contin})}(E) = \frac{S(E)\exp(-2\pi\eta)}{E}, 
\label{eq0}
\end{eqnarray}
where $\eta=\alpha Z_pZ_t \sqrt{\mu/2E}$, $\alpha$ = $e^2/4\pi\epsilon_o \hbar c$ $\approx$ 1/137 is the
fine structure constant.  For the case of a pure Coulomb repulsion,  the barrier penetrability is given by 
the Gamow-Sommerfeld penetrability
\begin{eqnarray}
P_{\rm contin}(E) = 2\mu \hbar^2 \exp(-2\pi\eta)/\pi,
\label{Pcont}
\end{eqnarray}
and the $S$-factor is $S(E)$ = 1. The reaction cross section $\sigma_R(E)$  
is then given by the Gamow-Sommerfeld cross section,
 \begin{eqnarray}
 \sigma_R^{({\rm Gamow-Sommerfeld})}(E) = \frac{\exp(-2\pi\eta)}{E}, 
 \end{eqnarray}
 shown as the dashed curve in Fig. 3(a) and (3(b). The energy dependencies of the  
reaction cross sections $\sigma_R$  from the potential scattering theory  are shown in Fig.\ \ref{fig3}(a).
 One can see the $\sigma_R$ of the potential scattering theory approximately equal 
to the Gamow-Sommerfeld  cross section at $E$ $\approx$  0.03  MeV and 0.8 MeV,
but is significantly enhanced over the Gamow-Sommerfeld cross sections near the resonance. 

To describe the scattering theory $\sigma_R$ results below the Hoyle resonance,  for $E~\! \le\!$ 0.26  MeV,
we can describe the effects of the interference of the resonance and  barrier penetration by  modifying  the $S(E)$ = 1 for Coulomb repulsion to  the modified Gamow reaction cross section with an energy-dependent  $S(E)$ factor given by 
\begin{eqnarray}
S(E) = \frac{1}{(s_0+s_1E+s_2E^2+s_3E^3)}
\label{MG}
\end{eqnarray}
 where $s_0$ = 1.28205 MeV$^{-1}$, $s_1$ = $-$11.763 MeV$^{-2}$, $s_2$ 
= 35.068 MeV$^{-3}$ and $s_3$ = $-$33.612 MeV$^{-4}$. Fig.\ \ref{fig3}(a) shows 
the modified Gamow cross section with Eqs.\  (17) and (\ref{MG}) matching the results of  
potential scattering theory in the continuum region.

Above the Hoyle resonance energy and 
 0.32  $< E_{\rm c.m.} <$ 1.77 MeV, where the $s$-wave contribution 
continues to be important  while other partial waves also contribute, 
we propose combining phenomenologically the Hill-Wheeler form of the barrier penetrability for the $s$ wave \cite{Hil53}  with the Wong expression for the sum over many partial waves \cite{Won73}, with fitting coefficients  $c_0$ and $c_1$ to represent the reaction cross section as
\begin{eqnarray}
P_{\rm contin}(E) &&=  \frac{c_0}{[1+\exp(2\pi(B_0-E)/\hbar\omega_0)]}
\nonumber\\
&& \hspace*{-0.5cm}+ c_1\ln\{1+\exp(2\pi(E-B_1)/\hbar\omega_1)\},
\label{HWW}     
\end{eqnarray}
where $c_0$ = 2.8$\times10^{-6}$, 
$c_1$ = 2$\times10^{-7}$, $B_0$ = 0.97 MeV, $B_1$ = 1.12 MeV, 
$\hbar\omega_0$ = 0.8607 MeV and $\hbar\omega_1$ = 0.2805 MeV, 
so that the subsequent Hill-Wheeler-Wong reaction cross section 
\begin{eqnarray}
\hspace*{-0.2cm} \sigma_R^{({\rm contin})}(E)&&~= \left(\frac{\pi}{k^2}\right) \Big(\frac{c_0}{[1+\exp(2\pi(B_0-E)/\hbar\omega_0)]}
\nonumber\\
&& \hspace*{-0.cm} +~c_1\ln\{1+\exp(2\pi(E-B_1)/\hbar\omega_1)\} \Big) 
\label{eq21}    
\end{eqnarray}
matches the results of potential scattering theory.

In Fig.\ \ref{fig3}(b), the solid curve shows the phenomenological penetrability $P(E, L=0)$ 
extracted from result of the potential scattering theory as a function of energy on a semi-log 
scale for the region near the Hoyle resonance. The penetrability in the Gamow model 
is shown as the dashed curve. The differences between the two curves indicate 
important interference between the resonance and the underlying potential 
barrier penetration.

\section{Comparison of resonances from potential scattering theory with experiment and 3$\alpha$-cluster HFE theory}
The collision between an $\alpha$ particle and a $^8$Be nucleus can be described by  the coupled-channel potential scattering theory, 
where the $\alpha$ particle scatters off the deformed potential of the $^8$Be nucleus. Resonances show up in the potential scattering, and they correspond to the resonances of the $^{12}$C compound nucleus. We seek a potential that will give  the proper resonances in agreement 
with  experimental $^{12}$C data.

In the previous sections, we showed that the constraint imposed by the  experimental 
$^{12}$C resonance energies and widths suggests 
 the need to include a parity-dependent surface potential component. By judiciously adjusting the potential parameters, which are  
listed in Table  \ref{tab0}, we obtain resonance energies ($E_\zeta$) and $\alpha$-widths ($\Gamma_\zeta$) in Table  \ref{tab1} 
for the $\alpha + ^8$Be collision at the center-of-mass energy $E_\zeta =E_{\rm c.m.}=E$, corresponding to the 
compound nucleus $^{12}$C$^*$ excited state with excitation energy $E_x = E_{\rm c.m.} + 7.367 $ MeV. 

\begin{table}[htp]
\centering
\caption{ The experimental $^{12}$C resonance energies $E_x$ 
and total widths $\Gamma_{\rm total}~= \Gamma_{\alpha} + \Gamma_{\gamma}$ 
of natural parity states, reported in Kelley, Purcell and Sheu \cite{Kel17}, John {\it et al.} \cite{John03},  Zimmerman {\it et al.} \cite{Zim13} 
and Freer {\it et al.} \cite{Fre11, Fre18}, compared with theoretical resonance energies $E_x$, $E_{\rm c.m.}$, 
and partial $\alpha$-width $\Gamma_\alpha$ in $\alpha$ +$^8$Be collisions 
obtained from the  coupled-channel potential scattering theory, and from the  3$\alpha$-cluster HFE theory \cite{AR07}. 
For displaying purposes, the theoretical $E_{\rm Hoyle}$ = 0.2809338737 MeV is truncated to $E$ = 0.281 MeV. 
The value in the () denotes to experimental uncertainty.} 
\vspace{0.3cm}
\resizebox{8.75cm}{!}{
\begin{tabular}{|c|c|c|c|c|c|c|c|}
\hline
State  &  \multicolumn{2}{c|}{Experiment}  &  \multicolumn{3}{c|}{Potential Scattering}  &  \multicolumn{2}{c|}{3$\alpha$-cluster }  \\
            &  \multicolumn{2}{c|}  {}                &  \multicolumn{3}{c|}{Coupled-Channel}  &  \multicolumn{2}{c|}{HFE Theory \cite{AR07}} \\
\hline
 $L^{\pi}$ &  $E_x$ & $\Gamma_{\rm  total}$ & $E_x$ & $E_{\rm c.m.}$ & $\Gamma_\alpha$ & $E_x$  & $\Gamma_\alpha$ \\  
                &  (MeV)  & (keV)                     & (MeV)  & (MeV)              &   (keV)  & (MeV)  & (keV)                     \\  \hline
0$_2^+$\cite{Kel17}   &7.65407(0.0019)  & 0.0093(0.0009)   & 7.648   & 0.281   & 0.008       & 7.63   & 0.0625  \\ 
0$_3^+$\cite{Kel17}  & 9.93     &2710     &10.159   &2.780    & 300        & 11.2   & 1000                    \\  
0$_3^+$\cite{John03}  & 10.3  &  3000   &              &              &            &                &                               \\      \hline  
1$_1^-$\cite{Kel17}  & 10.847(0.004)   & 273      &10.697     &  3.33    & 409         & 10.86   & 475        \\     
1$_2^-$                    &                          &           & 12.637    &5.270   & 610         &       &                           \\  \hline
2$_2^+$                    &                          &             &10.138     &  2.770   & 0.026       & 8.63    & 132        \\      
2$_2^+$\cite{Kel17}   & 9.87     & 850     &    &     &           &      &                                                                \\ 
2$_2^+$\cite{Zim13,Fre18}  & 10.03(0.05)    & 1600(130)   &    &                &                  &                 &              \\   
2$_3^+$\cite{John03} &  11.46 (0.2)   &  430 (100)    & 11.204  &   3.837  &    130       &  11.73  & 1086                                                \\ \hline
3$_1^-$\cite{Kel17}  &  9.641(0.005)   &  46      & 9.530    &2.165   & 38         & 9.58    &  68                   \\  
3$_2^-$                    &                          &           & 13.004    &5.637   & 95         &       &                               \\  \hline
4$_1^+$                &             &               &9.646     &2.279     & 12.4         & 10.5         &    396                \\   
4$_2^+$\cite{Fre11} & 13.3                & 1700      & 12.84     &5.473     & 1650         & 14.08         &    606   \\ 
4$_3^+$\cite{Kel17} & 14.079             & 272    &     &     &         &         &                   \\ 
\hline
\end{tabular}
}
\label{tab2}
\end{table}

\begin{figure}[htp]
\centering
\includegraphics[scale=0.34]{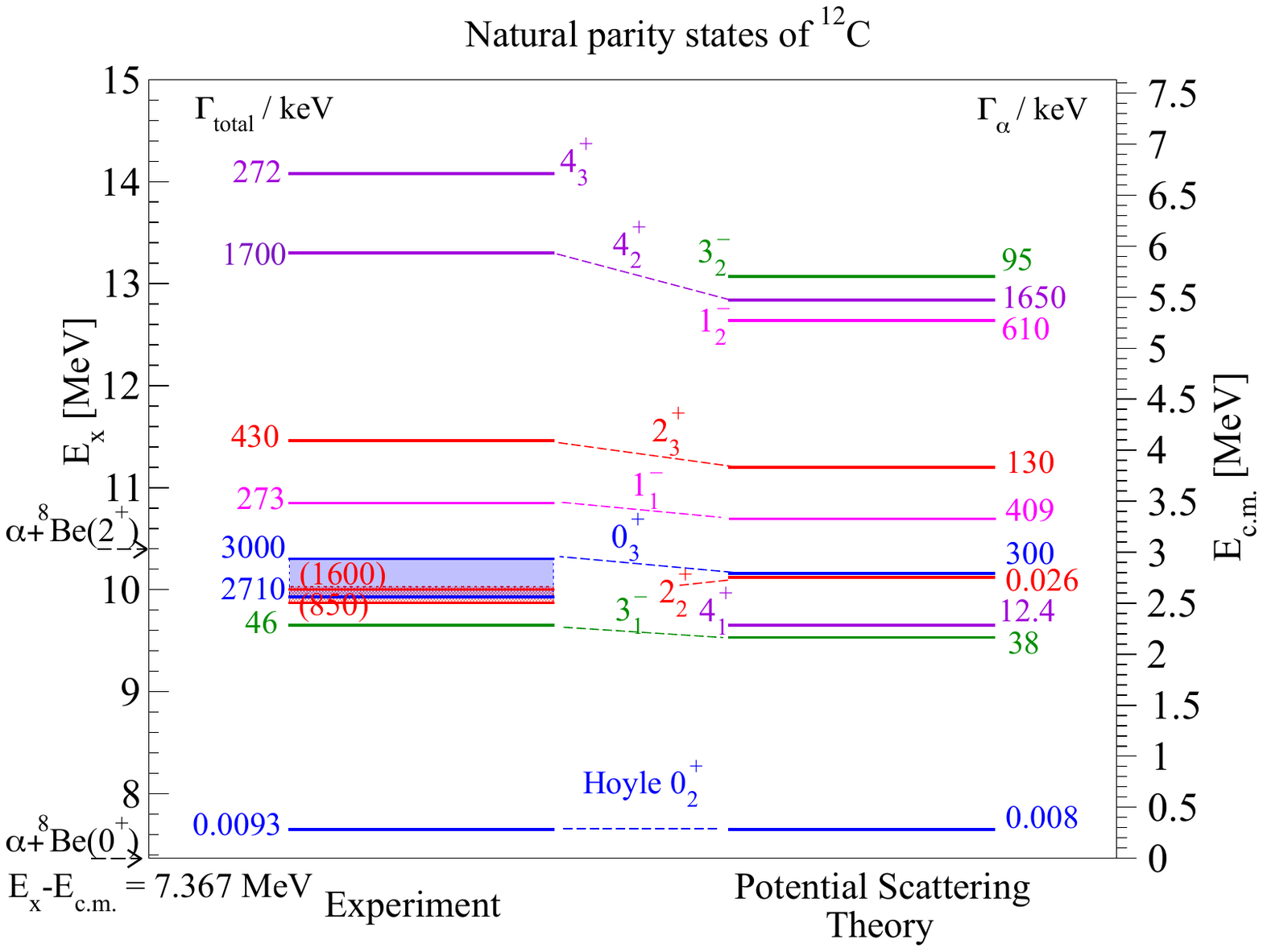}
\vspace{-0.7cm}
\caption{
Comparison of the experimental energies $E_x$  and total widths $\Gamma_{\rm total}$ 
of natural parity $^{12}$C resonances
with energies  and $\alpha$-widths $\Gamma_\alpha$ in $\alpha+^8$Be collisions obtained 
 from potential scattering theory coupled-channel calculations. 
The red band indicates the experimental 2$_2^+$ energy spreading from 9.87 to 10.03 MeV. 
The blue band indicates the experimental 0$_3^+$ energy spreading from 9.93 to 10.3 MeV.
}
\label{fig4}
\end{figure}

In Table \ref{tab2} and in Fig. \ref{fig4}, we compare the experimental $^{12}$C resonance energies ($E_x$) 
relative to the $^{12}$C ground state, and resonance total widths $\Gamma_{\rm total}$ from Ref. \cite{Kel17, John03, Zim13} 
with theoretical resonance energies and $\alpha$-decay widths ($\Gamma_\alpha$) obtained 
 from the coupled-channel potential scattering theory. 
In addition,  we also list in Table \ref{tab2} the theoretical results from
the $3\alpha$-cluster HFE theory  \cite{AR07} for comparison.

As the total widths, $\Gamma_{\rm total} =\Gamma_\alpha + \Gamma_\gamma$, of these low-lying resonances are dominated by the 
$\alpha$-width, $\Gamma_\alpha$, it is reasonable to compare the theoretical $\alpha$-width ($\Gamma_\alpha$) with 
the experimental total width ($\Gamma_{\rm total}$). Furthermore, in our comparison procedure, because the $\alpha$-width 
is a sensitive exponential function of the energy $E_x$ and we are not able to predict the resonance energy $E_x$ with a 
high degree of accuracy in this first  $\alpha+^8$Be study, it suffices to consider a comparison to be in approximate agreement 
if the energies  fall within 0.5 MeV, and the widths fall within the same order of magnitude, with their corresponding 
counterparts. 

The theoretical  energy of the $0_2^+$ resonance (Hoyle resonance) at $E_x = 7.648$ MeV and the $\alpha$-width of 8 eV agree with 
the experimental Hoyle resonance energy at 7.655 MeV and the total width of 9.3 eV. The theoretical resonance energy of the $1_1^-$ resonance  
at $E_x = 10.697$ MeV and the $\alpha$-width of 409 keV match the experimental $1^-$ resonance energy at 10.847 MeV and a total width 
of 273 keV \cite{Kel17}. The theoretical resonance energy of the $3_1^-$ resonance  at $E_x(3_1^-) = 9.530$ MeV and the $\alpha$-width 
of 38 keV agree with  the experimental $3_1^-$ resonance energy of 9.641 MeV and the total width of 46 keV. Finally, the 
theoretical resonance energy of the $0_3^+$ resonance  at  $E_x = 10.159$ MeV with a $\alpha$-width of 300 keV falls within 
the experimental $0_3^+$  resonance excitation energy of $E_x = 9.93 - 10.3$ MeV with a width of 2.71 MeV \cite{Kel17,John03}, 
within the same order of magnitude.

The comparison of the results from the potential scattering theory in the coupled-channel calculations
with those from the 3$\alpha$-cluster HFE theory  \cite{AR07} in Table \ref{tab2} shows approximate 
agreement in energies and widths, including the $4_1^+$ resonance.

For the $2_2^+$ resonance, even though the theoretical resonance energy at $E_x=10.138$ MeV closely matches 
the reported experimental $2^+$ resonance energy of 9.87 
\cite{Kel17}  and 10.03 MeV\cite{Fre18}, the theoretical $\alpha$-width of 26 eV differs 
from the experimental $\alpha$-width reported in Ref. \cite{Zim13,Fre18} by many orders of magnitude. 
On the other hand, this $2^+$ resonance at $E_x\approx 10$ MeV was not observed in the experiments of Ref. \cite{John03, Hyl09, Alco12, Kir12}.
These differences, along with the perplexing nature in the analysis of the $2_2^+$ resonance, will be discussed in Section 5.  
 
It is illuminating to examine the relationship between the resonance  energies and $\alpha$-widths 
of the $^{12}$C resonances, and their  effective potentials in Fig.\ 1. A pocket resonance can be identified by plotting the effective potential for the elastic channel 
of the  $L$-partial waves together with the resonance energy level, to determine whether the resonance at a 
given energy level is confined within the potential pocket. As illustrated in Fig.\ 1(a), 
the Hoyle ($0_2^+$) resonance at $E_x = 7.648$ MeV, the $1^-$ resonance at 10.847 MeV, and the $3^-$ resonance
at 9.530 MeV are located below their corresponding potential barriers. Consequently, 
these resonances can be classified as pocket resonances with narrow widths. Furthermore, the deeper the energy level of a resonance lies 
below the barrier, the narrower its $\alpha$-width.

For the $0_3^+$ resonance, experimental work by John {\it et al.} gave 
the experimental  $0_3^+$ resonance  energy at $E_x = 10.3$ MeV with a broad width of 2.71 MeV
\cite{John03,Kel17}.
In Table \ref{tab2}, the  scattering theory in coupled-channel calculations  gives the $0_3^+$ resonance energy 
at $E_x = 10.16$ MeV with a width of 300 keV, which yields approximate agreement  with the experimental energy 
level and approximate agreement within the same order of magnitude with the  width.  The theoretical $0_3^+$ resonance was also predicted  to be  $E_x = 10.3$ MeV with a width of 400 keV by using the  $3\alpha$-cluster antisymmetrized 
molecular dynamics ($3\alpha$-AMD) of Kanada-En'yo \cite{Enyo07}.

\begin{figure}[h]
\centering \includegraphics[scale=0.4]{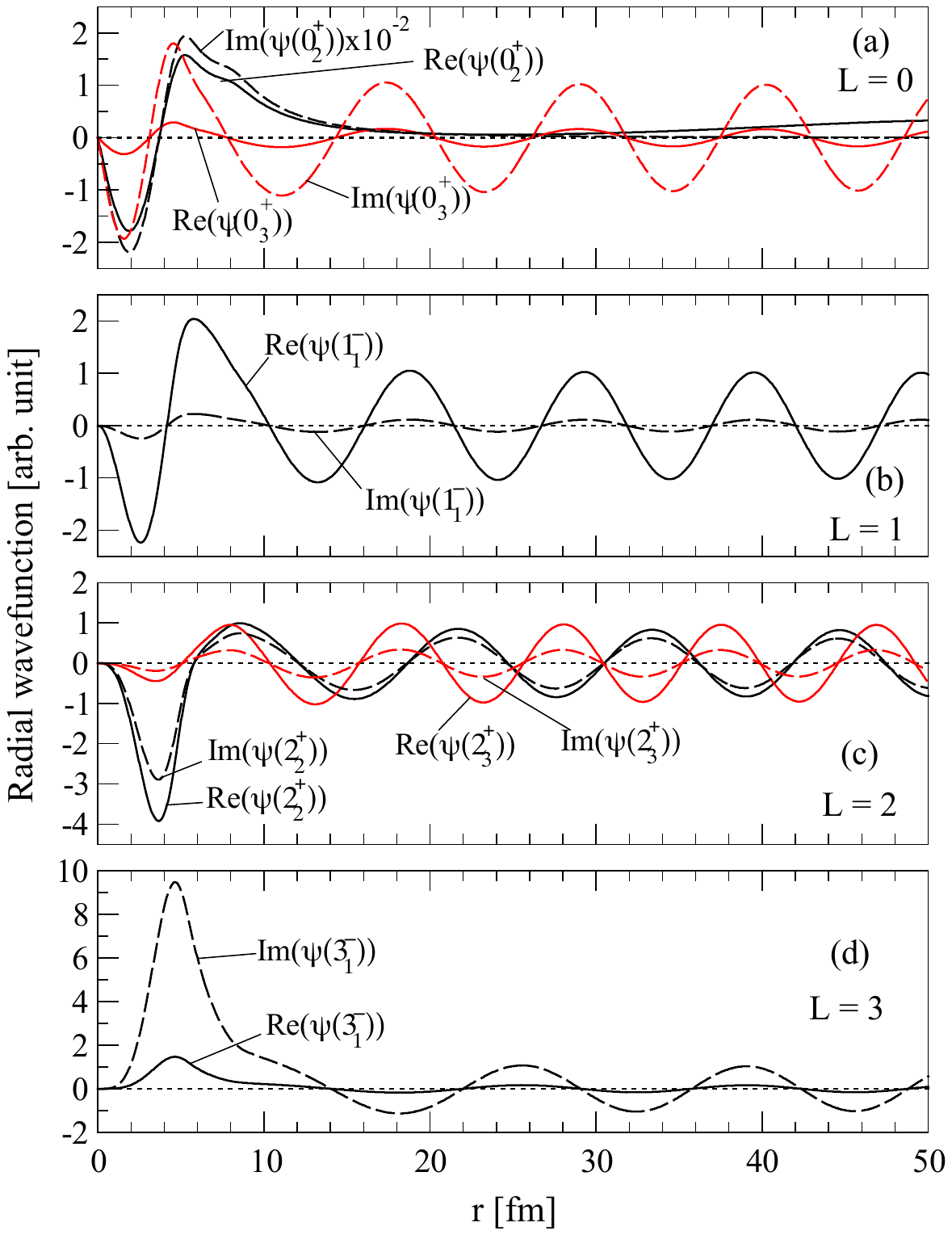}
\vspace{-1cm}
\caption{
Shown here are the real and imaginary parts of the radial wave functions for different  $L^{\pi}$ resonances  listed in Table \ref{tab2}.
}
\label{fig5}
\end{figure}

We can obtain useful insight into the nature of the resonances by studying the underlying potentials 
of different $L$-partial waves
in Fig.\ \ref{fig1}(a)  together with 
their corresponding  wave functions in Fig.\ \ref{fig5}. Because of the presence 
of parity-dependent surface potential  component as discussed earlier, the radial dependence of the total potentials for $^{12}$C $\{0^+ ,2^+ ,4^+ \}$ resonances
exhibits a double-hump structure and  possesses two local energy minima, corresponding to a doublet  of each of the $^{12}$C $\{0^+, 2^+, 4^+\}$ resonances, in the Hoyle resonance energy region.  The presence of the two 
 energy minima will enhance 
the wave function amplitudes around  the potential minima.  The lower-energy members of the doublet resonances
with the same $L^\pi$ quantum numbers need to tunnel through greater potential barriers and hence will have  narrower $\alpha$-widths
compared to their corresponding higher-energy counterparts. 

Accordingly, the lower-energy member of the $\{0^+\}$ doublet is the Hoyle $0_2^+$ resonance at $E_{\rm c.m.}=0.281$ MeV. 
As shown in Fig.\ref{fig5},
its probability density, $|\psi(0_2^+,r)|^2$, exhibits peaks at $r \approx 2.5$ and $5$ fm, predominantly occupying 
the interior region ($r<10$ fm) with only a small tunneling leakage into the exterior region (see Fig. 5(a)).
Due to its energy level  being well below the $L$ = 0 potential barrier, the Hoyle resonance has an extremely narrow width 
of approximately 8 eV. The corresponding wave function reveals pronounced tunneling behavior, characterized 
by a notable decaying amplitude within the potential well. In contrast, the higher-energy member of the $\{0^+\}$ doublet, the $0_3^+$ resonance  at $E_x = 10.16$ MeV, has a probability density $|\psi(0_3^+,r)|^2$ with peaks at $r \approx 2.5, 5, 10.5, 17,...$ fm. This resonance partially resides in the interior region but extends outward as an outgoing wave into the exterior region. Unlike the Hoyle $0_2^+$ resonance, the $0_3^+$ wave function exhibits only a weak tunneling behavior and features oscillations with nearly uniform internal and external amplitudes. The theoretical $\alpha$-width of the $0_3^+$ resonance  is significantly broader, approximately 300 keV.

Similarly, the lower-energy member of the $\{2^+\}$ resonance doublet at $E_x = 10.14$ MeV needs to tunnel through a greater potential barrier, and hence it has the narrower $\alpha$-width of 27 eV compared to its higher-energy counterpart, which has a width of 300 keV. The probability density $|\psi(2_2^+,r)|^2$ for this lower-energy $2_2^+$ resonance exhibits peaks at $r \approx 4, 8, 14, 20,...$ fm, with a slightly higher probability density in the interior region than in the exterior. In contrast, the upper member of the doublet, the $2_3^+$ resonance at $E_x = 11.20$ MeV, has probability density peaks at $r \approx 4, 8, 13, 18,...$ fm. Unlike the $2_2^+$ resonance, the $2_3^+$ wave function exhibits a lower peak magnitude in the interior region compared to the exterior. Its theoretical $\alpha$-width is 300 keV.

We must emphasize that while the lower-energy  $0_2^+$ member of the $\{0^+,2^+,4^+\}$ doublet may be identified as the Hoyle resonance, 
the analogous theoretically predicted lower-energy  $2_2^+$ narrow resonance at $E_x \approx 10.1$ has yet to be observed or identified. To test the
double-hump potential description with two sets of $\{0^+,2^+,4^+\}$ resonances, this resonance must be experimentally searched for and uncovered.
On the other hand, there exists an experimentally observed $2^+$ state at $E_x = 11.46$ MeV with a total width of $\Gamma_{\rm total} = 430 \pm 100$ keV \cite{John03}, along with the theoretically predicted $2_3^+$ resonance of the doublet member at $E_x = 11.20$ MeV with an $\alpha$-width of 130 keV. Because of the similarity in resonance energies and widths, it is reasonable to identify this theoretical higher-energy $2_3^+$ resonance with the experimental $2^+$ resonance observed in Ref.\cite{John03}.

The  lower energy $4_1^+$ resonance, the analog of the Hoyle resonance with angular momentum $L = 4$, is theoretically predicted to lie at $E_x = 9.6$ MeV with a width of 12 keV. However, this resonance has not yet been observed or identified, and an experimental search is required to test the double-hump potential description.
On the other hand, the theoretically predicted $4_2^+$ resonance of the $4^+$ doublet at $E_x = 12.84$ MeV has an $\alpha$-width of 1650 keV.  There is experimentally an observed $4^+$ resonance at $E_x = 13.3$ MeV with a total width of $\Gamma_{\rm total} = 1700$ keV \cite{Fre11}. Again,  the similarity in resonance energies and widths suggest it reasonable to identify the theoretical  the higher-energy member of the  of the $4^+$ resonance doublet with the experimentally observed $4^+$ resonance reported in Ref.\cite{Fre11}.

For the $3_1^-$ resonance, its energy and width show remarkable agreement with both experimental data and the $3\alpha$-cluster HFE theory. Notably, recent experiments suggest that the $3_1^-$ resonance may enhance the $3\alpha$ reaction rate \cite{Tsu21}.
Furthermore, our predictions include two $4^+$ resonances.  The first, $4_1^+$ resonance, is in close proximity to the $3_1^-$ resonance but at a slightly lower energy than suggested by the 3$\alpha$-cluster HFE theory. The second $4_2^+$ resonance appears to correspond to the second $4_2^+$ 
resonance of the 3$\alpha$-cluster HFE theory, which agrees with the experimental observations at $E_x = 13.3$ MeV.

It is interesting to note the similarity of the double-hump feature of the effective potentials for the $\{0^+,2^+,4^+\}$ resonances  in the present
work with  those in  fission isomers  \cite{Bra72} and  in 
shape isomerism of Hg isotopes \cite{Kol75}
where the double potential  well gives rise to a doublet of resonance states of the same quantum numbers centering  around the two different potential minima.
 The  double-hump potential in $^{12}$C potential results in the occurrence of doublet of  $\{ 0^+, 2^+,4^+\}$ resonances with  the lower-energy members of these resonance states possessing a narrower width compared to the higher-energy members, 
as they must tunnel through barriers of greater heights.  The 3$\alpha$-cluster HFE theory with parity-dependent three-body forces, also gave these 
doublet  resonances with the same angular momentum. The prediction of such doublet resonance states in the Hoyle resonance energy region will be an 
interesting experimental signature  for the double-hump barrier for even parity states in $^{12}$C.

It is necessary to point out that  in addition to   the natural-parity states we have considered, the  experimental $^{12}$C spectrum contains unnatural parity states. 
The potential scattering theory we have considered involves  a scalar potential  which can only give rise to resonances of natural parity, with (parity $\pi$)=$(-1)^L$.  Unnatural parity  resonances with $\pi=(-1)^{(L+1)}$ will involve a potential of pseudoscalar nature and is beyond the scope of the present consideration. 

\section{Questions on the $4_1^+$ and $2_2^+$ resonances of  $^{12}$C at $E_x \approx$ 10 M\lowercase{e}V}
The comparisons in Fig.\ \ref{fig4} in the last section indicate that 
the potential scattering theory gives resonance energies and widths  approximately consistent with 
 the experimental $^{12}$C data.   That is, the theoretical and experimental counterparts agree within half an MeV in resonance energies  and
 within the same order of magnitude  in resonance widths. 
  However, the theoretically predicted  
lower energy  $4_1^+$ and  $2_2^+$ resonances at $E_x \approx 10$ MeV are yet to be identified or observed.
There may be many possible reasons which  could contribute to the difficulties in identifying
or observing these predicted resonances.  
We shall  discuss some of  these difficulties in 
observing the narrow $4^+$ and $2^+$ resonances at $E_x\approx 10$ MeV.
We shall also present a  theoretical  analysis of the potential  landscape  for the $2^+$ resonance which governs the tunneling of $\alpha$ and the $\alpha$-width at $E_x\approx 10$ MeV.

It is  theoretically significant that the lower energy $2_2^+$ and  4$_1^+$ resonances at $E_x\approx 10$ MeV have been  predicted not only by the present
scattering theory, but also by  the
3$\alpha$-cluster HFE model of Jensen and collaborators at
$E_x=$ 8.655 MeV with an $\alpha$-width of 132 keV for the $2_2^+$ resonance, and 
 $E_x=$ 10.525 MeV with an $\alpha$-width of 396 keV for the 4$_1^+$ resonance \cite{AR07}.  Furthermore, 
they have also been  predicted by a third theoretical 3$\alpha$-AMD model of
Kanada {\it et al.}  \cite{Enyo19}  at $E_x=$  10.6 for  the $2_2^+$ resonance and 10.9 MeV for the 4$_1^+$ resonance.
The common predictions of the $2_2^+$ and $4_1^+$ resonances at $E_x\approx 10$ MeV by three different theoretical models suggest that  the possible existence of the $2_2^+$ and  $4_1^+$  resonances 
at $E_x\approx 10$ MeV  should be taken seriously.   
Therefore, future experimental  search for the $2_2^+$ and $4_1^+$ resonances at $E_x\approx 10$ MeV would be of interest.

\subsection{Difficulty in the observation of the $4^+$ resonance at $E_x\approx$ 10  MeV}
A possible explanation for the as-yet unobserved $4^+$ resonance is that the $4^+$ resonance predicted 
at 9.646 MeV, with a width of approximately 10 keV, is located nearly at the same energy as the $3^-$ resonance predicted
 at 9.641 MeV with an $\alpha$-width of about 50 keV. As a consequence,  
the weaker excitation probability of the higher multipole 4$^+$ resonance relative to
the 3$^-$ resonance in a multipole interaction may make the lower $4^+$ resonance hidden under the 3$^-$ resonance. 
Furthermore, the widths of these resonances are sensitive to the resonance energy level relative to the top of potential barrier and the potential curvature \cite{Won73}. Consequently, there is a high degree of uncertainties in the predicted width. 
The ability to observe $-$ or the failure to observe $-$ the low-energy $4^+$ resonance could provide 
valuable insights into the nature of the barrier the compound $4^+$ system must tunnel through.
It will therefore be of great interest to explore future angular correlation measurements of high
precision, high intensity, and fine binning such as those carried out in Freer {\it et al.} \cite{Fre12}
to see whether the possible $4^+$ resonance can be separated and distinguished from the observed 
broad $3^-$ resonance. Previously, the possibility that the  $4^+$ resonance at $E_x \approx 10 $ MeV   could be hidden in other resonances was already noted in Ref.\cite{AR07}. 

\subsection{The perplexing $2^+$ resonance at $E_x\approx 10$ MeV} 
For the $2^+$ resonance
predicted to locate at 
$E_x = 10.138$ MeV with a width of 27 eV,  
a complication arises because the narrow $2_2^+$  resonance nearly coincides with   the 
  $0_3^+$ resonance at $E_x= 10.159$ MeV with a width of 300 keV, as shown in Fig.\ 1(b).  The close proximity of the $2_2^+$ and $0_3^+$ resonances 
and the narrow width of the $2_2^+$ resonance  may make it difficult to separate them. There is the 
possibility that the narrow $2_2^+$ resonance  at $E_x\approx 10$ MeV may be  
hidden under the broad $0_3^+$ peak experimentally. 

\begin{figure}[htp]
\centering
\includegraphics[width=1.0\linewidth,height=0.65\linewidth]{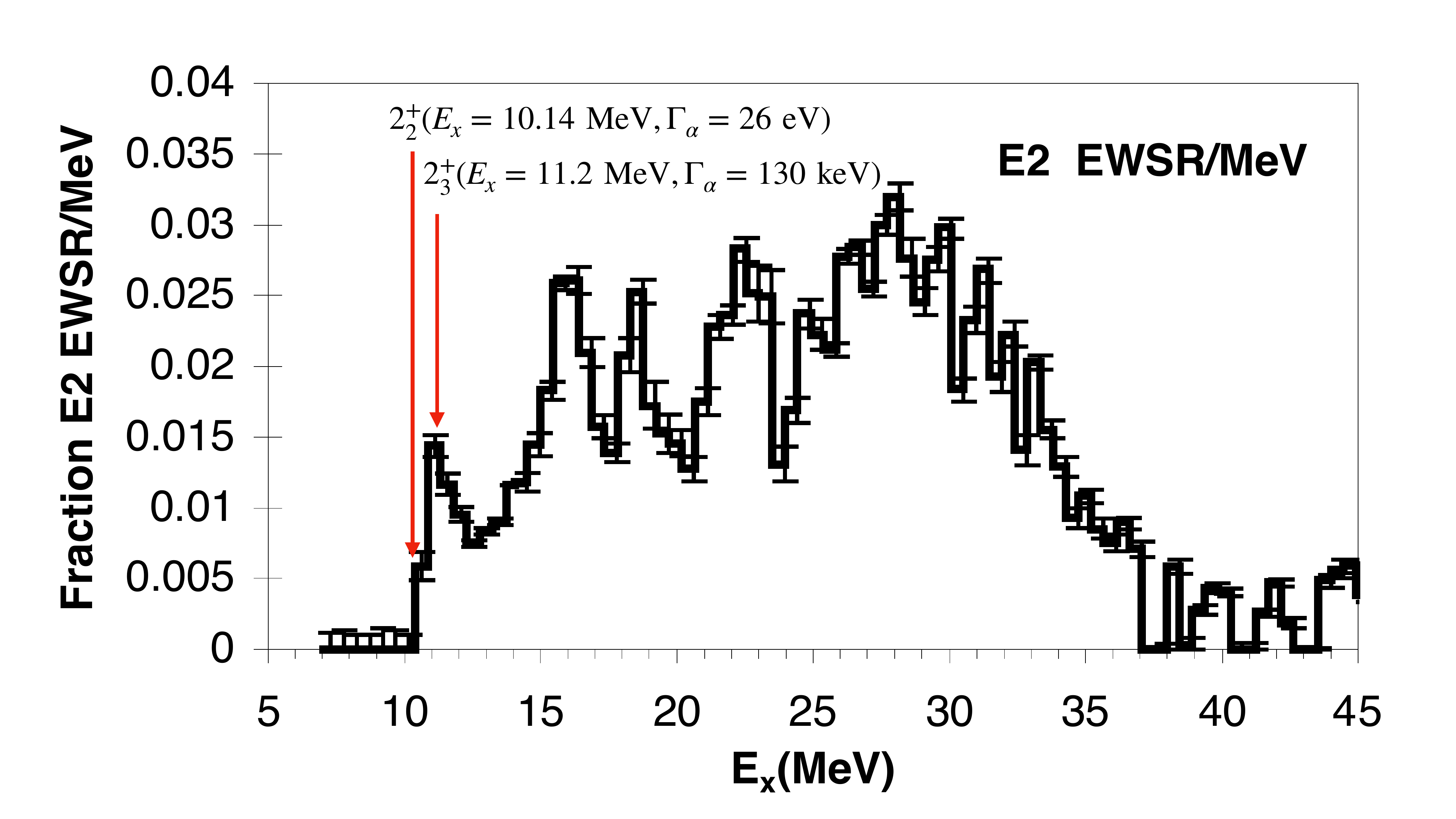}
\vspace{-0.8cm}
\caption{
The experimental isoscalar $E2$ strength distribution
data as shown in Figure\ 11(c) of John {\it et al.} \cite{John03}.
The experimental data were  obtained from the multipole excitation fits to the 
 $^{12}$C ($\alpha,\alpha')$ fit to the inelastic differential  cross sections at $E_\alpha$=240 MeV.  The locations of the $2_2^+$ and $2_3^+$  resonances predicted by the scattering theory coupled-channel calculations are marked by arrows.
 }
\label{fig11c}
\end{figure}

The experimental presence or absence of the $^{12}$C $2^+$ resonance  at $E_x\approx 10$ MeV may also be perplexingly uncertain 
because different experiments gave different results. 
The inelastic scattering experiment by John {\it et al.} measured the angular distributions 
of $\alpha$-particle on $^{12}$C at $E_\alpha$ = 240 MeV \cite{John03}. 
The angular distribution data were 
fitted to the results from the distorted wave Born approximation (DWBA) calculations. 
From the $2^+$ strength function measurements as shown in Fig.\ 6, 
John {\it et al.}  \cite{John03} observed 
a pronounced  2$^+$ peak 
in 
the experimental $2^+$
at  the excitation energy $E_x = 11.46 \pm
0.2$ MeV with a  width of $0.43 $ MeV.
However, they 
 did not observe a $2^+$ resonance at $E_x \approx 10$ MeV. 
Instead, the strength function  exhibits a sudden rise  at $E_x\approx $ 10 MeV, as shown in Fig.\ \ref{fig11c}, where the locations of the two  
theoretical $2^+$  resonances from the CC calculations are also indicated by arrows. It will be of interest to examine in future 
experiments with fine energy resolutions whether the sudden rise in the strength function may suggest the possible presence 
of a narrow $2^+$ resonance at $E_x \approx 10$ MeV.   On the other hand, the  pronounced  2$^+$ peak observed by John {\it et al.} at  the higher excitation energy $E_x = 11.46 \pm
0.2$ MeV and a  width of $0.43 $ MeV may correspond to the theoretically predicted $2_3^+$ resonance
at $E_x = 11.2$ MeV with an $\alpha$-width of 0.13 MeV, as discussed in the last section (see Table III). 

In the $\beta$-decay experiment of $^{12}$N and $^{12}$B to study the $^{12}$C resonances, 
Hyldegaard {\it et al.} \cite{Hyl09} sought to 
resolve the interfering $^{12}$C $0^+$ and $2^+$ strengths 
in the 3$\alpha$ continuum by considering all major breakup channels.
The breakups can 
proceed either through the  $^8$Be ground state or the $^8$Be excited $2^+$ resonance
involve different angular correlations. Multichannel $R$-matrix method incorporating interfering resonances and several decay
channels was used to analyze the breakup involving different angular correlations. The $R$-matrix fit 
to the branching ratio as a function of 3$\alpha$ energy indicates 
the presence of the Hoyle resonance at 7.65 MeV,  the $0_3^+$ resonance at 11.2 MeV with a width 
of 1.5 MeV, and  the $2_2^+$ resonance at 11.1 MeV with a width of 1.4 MeV.
 However, there was no indication of a $2^+$ resonance at $E_x~\approx$ 10 MeV in the beta decay measurement of Ref. \cite{Kir12}.

 In another experiments based on $^3$He induced reactions 
 $^{10}$B($^3$He, $p \alpha\alpha\alpha$) at 4.9 MeV 
 and $^{11}$B($^3$He, $d \alpha\alpha\alpha$) reactions at 8.5 MeV,
  Alcorta {\it et al.}  \cite{Alco12} and Kirsebom {\it et al.}  \cite{Kir12}
studied the resonances of $^{12}$C$^*$ in transfer reactions.
They used  
the transfer reaction $^3$He + $^{10}$B $\rightarrow$ $p~+~^{12}$C* 
and $^3$He + $^{11}$B $\rightarrow$ $d~+~^{12}$C*
to  populate   $^{12}$C$^*$   resonances up to an excitation energy of 15 MeV. The $^{12}$C$^*$ 
 excitation spectrum could be separated into decays that proceeded via the ground state or excited states of $^8$Be by measurements in complete kinematics. 
 Six distinct excited resonances were observed: the Hoyle resonance at 7.65 MeV, the $3^-$ resonance at 9.64 MeV, the $1^-$ at 10.84 MeV, $2^-$ at 11.83 MeV, $1^+$  at
12.71 MeV, and finally the  $4^-$ resonance at 13.35 MeV. However, there was no indication  of a $2^+$ resonance at $E_x~\approx$10 MeV. 

On other hand, many other experiments reported the presence of a  $^{12}$C $2^+$ resonance  at $E_x~\approx$ 10 MeV with a broad width, of the order an MeV.
Inelastic scattering experiments on $^{12}$C($p,p’$)
and $^{12}$C($\alpha, \alpha’$) reactions conducted by Freer and coworkers
\cite{Itoh11, Fre12} utilized single-channel $R$-matrix
fitting to analyze the 2$^+$ line shape in the $^{12}$C excitation energy
spectra. Their analysis suggested a width of 750(150) keV centered at $E_x$ =
9.75(0.15) MeV. However,  the presence of the prominent $3_1^-$ resonance at 9.6 MeV 
and/or the $0_3^+$ resonance at 10.3 MeV may obstruct the precise extraction of the 
narrow $2_2^+$ resonance spectrum.

Recently, Kanada-En'yo and Ogata analyzed the $^{12}$C($\alpha,
\alpha'$) inelastic angular distribution using 3$\alpha$-AMD method and 3$\alpha$-AMD with
generator coordinate method (GCM) within a
coupled-channel approach \cite{Enyo19}.  Their analysis showed that in
the region of $E_x~\approx$ 10 MeV the experimental
$^{12}$C$(\alpha,\alpha')$ inelastic angular distribution at
$E_\alpha$ = 386 MeV could be understood as an incoherent admixture
from a $2_2^+$ resonance at 9.84 MeV and a $0_3^+$ resonance at 9.93 MeV \cite{Itoh11}. However,
the experimental inelastic scattering angular distribution data at
$E_\alpha$ = 240 MeV of Ref. \cite{John03} disagreed with their
theoretical angular distribution for such a $2_2^+$ resonance and $3_1^-$ resonance
admixture, where a $2_2^+$ resonance angular distribution peak at
$\theta~\approx 5^o$ was predicted in 3$\alpha$-AMD and 3$\alpha$-AMD+GCM calculations,
whereas the experimental data indicated a valley distribution
\cite{John03} (see Fig.\ 4 of Ref. \cite{Enyo19}).  The disagreement raises
questions about the consistency of the experimental suggestions of the
strengths of the $2_2^+$ and $3_1^-$ admixture in the
$^{12}$C$(\alpha$, $\alpha'$) analysis of \cite{Enyo19}.  This example
illustrates the difficulties and uncertainties  in the study of the $^{12}$C
$2^+$ resonance at $E_x~\approx$ 10 MeV.

Zimmerman {\it et al.} \cite{Zim13} carried out 
a high-resolution photodisintegration 
experiment to measure the differential cross section 
$d\sigma(E_\gamma)/d\Omega_{\gamma \alpha}=W(\theta_{\alpha \gamma}, E_\gamma)$ for  the $^{12}$C($\gamma$,$\alpha$)$^8$Be reaction 
within the excitation energy range $E_x({}^{12}$C$^*)$ from  9.1 to 10.7 MeV. 
The admixture of the electromagnetic $E1$ and $E2$ multipole radiations 
in the photodisintegration process  was then extracted 
 from the measured $W(\theta_{\alpha \gamma}, E_\gamma)$,
  as a 
function of the excitation energy $E_x({}^{12}$C$^*$).
They reported the observation of a broad $2^+$ resonance at 10.03 MeV 
with a width in the range of 0.8 to 1.6 MeV.  

In Ref. \cite{Zim13} and Zimmerman's thesis \cite{Zim13a}, the derivation of the underlying theory foundation
for the extraction of the width was not presented in detail.  We shall attempt to present  a derivation 
and  suggest the need to include the important $^{12}$C deformation effect in future analysis.    

If one wishes to carry out a theoretical analysis of the photodisintegration experimental data along the 
lines of Ref. \cite{Zim13, Zim13a}, 
one begins with the reciprocity theorem of nuclear reactions \cite{Bla52},
\begin{eqnarray}
&& \!\!\frac{ d\sigma( \gamma \!+\! ^{12}{\rm C} \!\to ^{12}\!\!{\rm C}^*\!\to 
 \!\alpha \!+\! ^8{\rm Be})    }{d\Omega_{\gamma\alpha} (\theta_{\gamma\alpha}) }
\!
\nonumber\\
&&=\left  (\frac{k_{(\alpha {}^8{\rm Be})}^2}{k_{(\gamma {}^{12}{\rm C)} }^2} \right ) 
\!\!\frac{ d\sigma(  \alpha \!+\! ^8{\rm Be}\!\to^{12}\!\!{\rm C}^*\!\!\to \! 
\gamma \!+ \!^{12}{\rm C} )}{d\Omega_{\alpha\gamma} (\theta_{\alpha\gamma}) },
\label{dcs}
\end{eqnarray}
where 
\begin{eqnarray}
k_{(\gamma  {}^{12}{\rm C)} }&=& \sqrt{E_\gamma}/\hbar , \\
k_{(\alpha {}^8{\rm Be})}& =& \sqrt{2\mu (E_\gamma - Q)}/\hbar, 
\end{eqnarray}
$\mu$ = 2483.6 MeV/$c^2$, and
  $Q$ = $M({}^{8}{\rm Be})c^2$+$M(\alpha)c^2$-$M({}^{12}{\rm C})c^2$ 
  = 7.367 MeV.
Upon taking the photon angular distributions in the electromagnetic decay of 
$^{12}$C$^*$ (on the right hand side of Eq.\ (\ref{dcs}) as 
a function of the multipole admixtures  at a given excitation energy $E_x$
as presented in  Blatt and Weisskopf \cite{Bla52}, and assuming that the polar axis of $^{12}$C$^*$ is along the alpha particle direction,  
then the experimentally measured angular distribution $W(\theta_{\alpha \gamma}, E_\gamma)$
 in the photo-excitation of $^{12}$C from 
the ground state to the excited $^{12}$C$^*$ state at a $E_x ({}^{12}$C$^*)=E_\gamma$ 
(on the left-hand side of Eq.\ (\ref{dcs})) can be used to determine the multipole admixtures 
as a function of excitation energy  $E_x$.
Zimmerman and collaborators \cite{Zim13, Zim13a} obtained  a broad  $2^+$ resonance of $^{12}$C at 10.03 MeV with a width of 1.6 MeV, 
admixed with a $1^-$ resonance at $E_x=10.84$ MeV with a width of 273 keV. However, the application of the angular distribution for the EM decay of a spherical nucleus to the photo-excitation or de-excitation of $^{12}$C may be subjected to questions.
The assumption that the polar axis of $^{12}$C$^*$ is along the alpha particle direction may also be questioned. 
Significant nuclear deformation effects need to be included in the photo-excitation 
and the $\alpha$-emission analysis before the extracted results of Ref. \cite{Zim13} can be considered definitive.

To study properly the angular correlation
between $\gamma$ and $\alpha$ in the $^{12}$C$(\gamma$,$\alpha)^8$Be$_{\rm gs}$ reaction of Ref. \cite{Zim13}, one needs to carry out first  the  photo-absorption of an $E1$ or $E2$ photon by $^{12}$C.
The polarization of the absorbed $E1$ or $E2$ photon is perpendicular to the direction of the photon propagation.  
For the photo-excitation  to the produced  $^{12}$C$^*$ state, the spin and the matrix element 
depend on the photon polarization.  As a consequence, 
the compound nucleus resonance state $^{12}$C$^*$ has  spin aligned along the direction of the photon propagation.   The photon-absorption  angular distribution  will depend on 
the opening angle between the direction of the absorbed photon and the intrinsic symmetry axis  as well as  on the deformation of the $^{12}$C nucleus.  Subsequent emission of the $\alpha$ particle from the produced  $^{12}$C$^*$ $1^-$ or $2^+$ depends on the tunneling of the $\alpha$ particle through the potential barrier. 
 Depending on the orientation  and the  deformation of the $^{12}$C$^*$ nucleus, different $\alpha$
 escape paths will lead to different degrees of potential barrier penetration, leading to  another source of anisotropy for the angular distribution  between $\alpha$ and $\gamma$.  The quadrupole deformation of the $^{12}$C nucleus will have a  significant effect on the quadrupole anisotropy of the $\theta_{\gamma\alpha}$ angular distribution. 
 The angular correlation of $\gamma$ and $\alpha$ considered here in 
 the sequence of reactions of $^{12}$C($0^+ )$$\xrightarrow{\gamma}$$^{12}$C$^*$$( 1^- {\rm or}~ 2^+)$$ \xrightarrow{\alpha}^{8}$Be$(0^+)$ may be mathematically considered as a special case of the angular correlation 
 between $\gamma$ and another $x$ particle, 
 for $x=\gamma,n,p,\alpha$,  
 in the  decay of a prolate or oblate nucleus along the sequence of 
 $I_1$$\xrightarrow{\gamma}I_2\xrightarrow{x}I_3$.
 There, in the latter case, the angular correlation between $\gamma$ and $x$  contains an additional nuclear deformation factor which in general can be expressed
   in terms of the $A_2$ and $A_4$ coefficients 
   as in Ref. \cite{Yam67,Der74,Wad98}, 
\begin{eqnarray}
w_{\gamma x}(\theta)=1+ A_2 P_2(\cos \theta)+A_4 P_4(\cos \theta),
\end{eqnarray}
 where $P_K$ are the Legendre polynomials.
The coefficient $A_2$ of $P_2$ have been found to depend significantly on the deformation of the nuclei in this sequence of particle correlations  \cite{Yam67,Der74,Wad98}.
  Therefore, if we treat $^{12}$C as a spherical nucleus, then the $A_2 P_2(\cos \theta)$ contribution from the $^{12}$C deformation may be distributed to arising from an  electric quadruple $E2$ source.   It is therefore important to take into account the $^{12}$C  deformation effect in the angular correlation of the $\alpha$ and the $\gamma$ photon in the analysis of photodisintegration of $^{12}$C.
Neglecting the intrinsic deformation  effects by treating the oblate $^{12}$C as spherical, in the photodisintegration of $^{12}$C, as done in the work 
of Ref. \cite{Zim13}, may likely introduce large consequential uncertainties.

There is another important effect for the $E1$ photo-excitation of $^{12}$C which will need to be considered further.  The distribution of the $1^-$ strength as a result of  the direct $E1$ photo-excitation was considered to follow
a single Breit-Wigner energy distribution in Ref. \cite{Zim13}. However, the direct $E1$ excitation of the $^{12}$C ground state leading to the $1_1^-$ resonance or continuum state depends on the geometrical shape of the $^{12}$C ground state.  The oblate spheroidal (pancake-like) shape of $^{12}$C possesses two length scales with major and minor axes. This could lead to the splitting of dipole strengths into two components with different energies, similar to the broadening or splitting of giant dipole resonances in deformed nuclei \cite{Spi69}. The representation of such $E1$ excitation strengths over the broad energy region from $9.1 < E_\gamma < 10.7$ MeV should also consider the oblate deformation as a significant spreading factor in the energy distribution of the $1_1^-$ resonance. Clearly, this shape of the $E1$ strength distribution will significantly affect the separation of experimental data into $E1$ and $E2$ components and the $\alpha$-width of the $2^+$ resonance.

In view of the above, the broad $2_2^+$ resonance of $^{12}$C at $E_x~\approx$ 10 MeV,
as reported in Ref.\cite{Zim13}, may not yet be definitive, and require further 
experimental and theoretical investigations on the effects the  $^{12}$C deformation.

\subsection{The $\alpha$-width $\Gamma_\alpha$ and the potential landscape for 
$\alpha$ 
 tunneling for  the $2_2^+$ resonance
at $E_x~\approx$ 10 MeV}
The $\alpha$-width $\Gamma_\alpha$ of a  $^{12}$C resonance is a physical quantity. 
At the resonance energy,
 it  gives the rate of escape (or tunneling) for 
 the $\alpha$ particle  from the interior region to the outside. 
The magnitude of $\Gamma_\alpha$ depends sensitively on the relative difference between  the $\alpha$-particle 
energy level and the top of the potential barrier \cite{Hil53}. Therefore, valuable information on the $\alpha$-width $\Gamma_\alpha$ of the $2^+$ resonance at $E_x~\approx$ 10 MeV can be obtained by studying 
 its energy level relative to the the potential  landscape.

Before we study the $\alpha$-width of the  the $2_2^+$ resonance at $E_x$ $\approx$ 10 MeV in question, 
we first examine other simple cases to gain a general idea on the magnitudes of the $\alpha$-widths 
and its relation to their potential barriers. In the $\alpha$+$^8$Be potential depicted in Fig.\ 1(a), 
the Hoyle $0_2^+$ resonance has an energy level well below the potential barrier.   
Consequently, the Hoyle $0_2^+$ resonance
is a pocket resonance with  a very narrow $\alpha$-width, on the order eV.  The
$1_1^-$ and $3_1^-$ pocket resonances, situated at higher energy levels have less
obstructing potential barriers. As a result, they have greater probabilities 
for penetrating the barrier, leading to their $\Gamma_\alpha$ increase to the keV range. 
For these resonances with energy levels below the barrier, their widths remain well under around 1.0 MeV range.

However, if the system energy level is such that its energy level lies near or
considerably above the top of the potential barrier, then its $\Gamma_\alpha$ will be large. According to the CC calculations, 
an $\Gamma_\alpha$ on the order of MeV is indicative of the system energy level being substantially 
above the potential barrier. 

\begin{figure}[htp]
\centering
\includegraphics[width=0.48\textwidth]{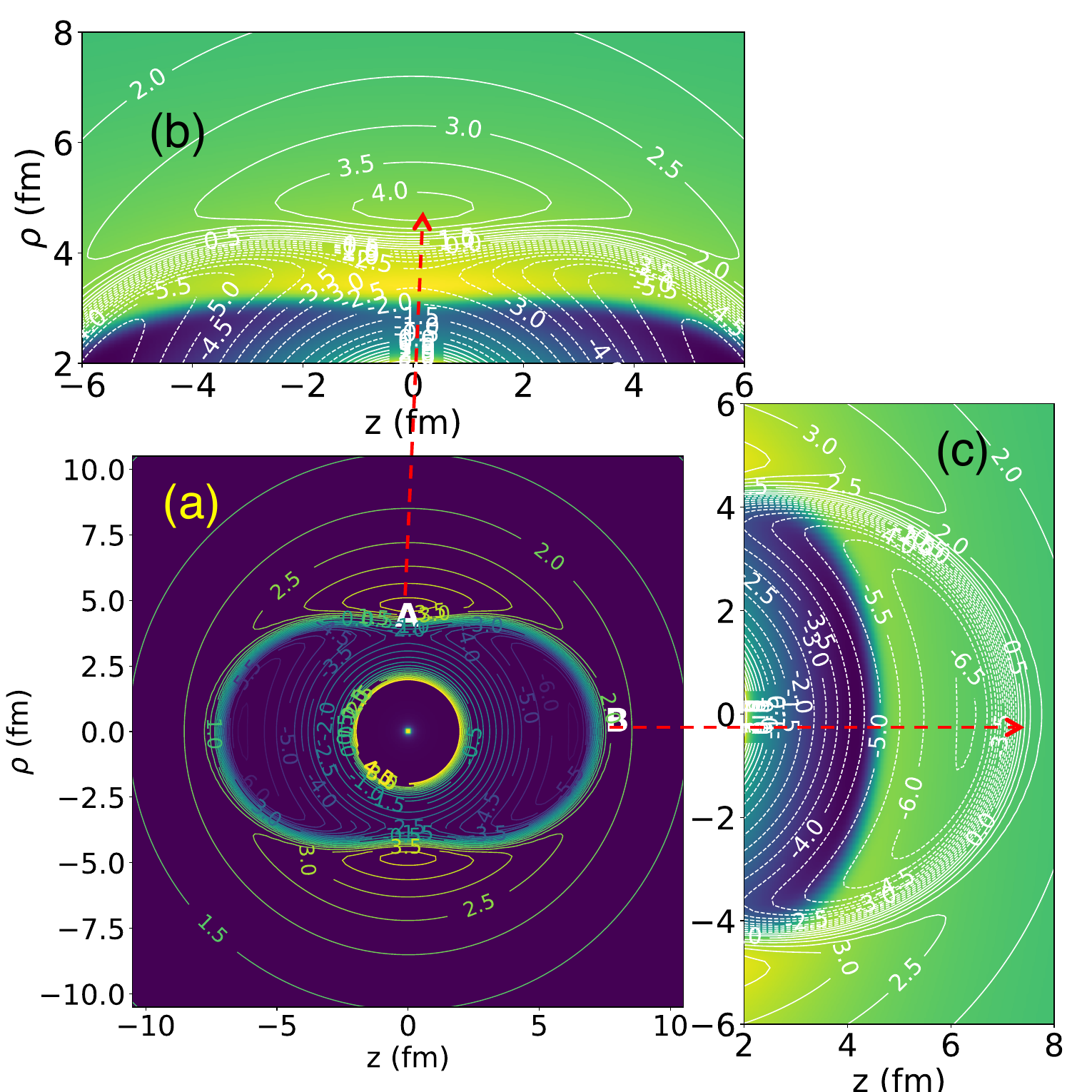}
\vspace{-0.7cm}
\caption{
(a) Contours of potential energy surface in cylindrical coordinates, for  $\alpha$-particle interacting with the deformed $^{8}$Be nucleus in the
for $L$ = 2 orbital angular 
momentum state as given by 
Eq.\ (1).  (b) Plot zooms into Zone A. (c) Plot zooms into  Zone B.
}
\label{fig_CS}
\end{figure}

To see whether the $\alpha$-width 
of the $2_2^+$ resonance at $E_x~\approx$ 10 MeV in question 
belongs to the below-the-barrier category 
or the above-the-barrier category, we can compare its
energy level at $E_{\rm c.m.}~\approx$ 2.6 MeV (or $E_x~\approx$ 9.7 MeV)  
of the compound system relative to the top of  the potential landscape,  in an approach
similar to Hill's three-body problem analysis by Wheeler \cite{Whe63}.   

The interaction potential of Eq.\ (\ref{Vtot}) between $\alpha$ and
$^8$Be for $L=2$, which includes the 
the $^8$Be deformation 
effects as well as 
Coulomb and centrifugal barriers, gives a potential landscape as depicted in Fig. \ref{fig_CS}.
We observe a potential barrier peaking at approximately $E~\approx$ 4 MeV in zone A in the transverse direction of the prolate
$^8$Be$_{\rm gs}$ nucleus, and a barrier of about $E~\approx$ 2.6 MeV
along the longitudinal direction alongside a saddle point in zone
B. Figures \ref{fig_CS}(b) and (c) provide detailed views of these
regions.  

We consider a compound $2^+$ resonance of excited $^{12}$C$^*$ system 
produced by the direct absorption of a
$E2$ photon at the $2_2^+$ resonance energy at $E_x~\approx$ 10 MeV as reported 
in the experiment of \cite{Zim13}.  At the  energy level of $E_x~\approx$ 10 MeV, the system now resides in 
the ``skating-race-track-like'' potential configuration of Fig. 7 which features
potential walls that can trap the incoming $\alpha$-particle wave.  
Once captured through a narrow passage near zone B, the system trajectory
may ricochet within the confines of the elevated barriers at other
orientations, transiently forming the meta-stable $^{12}$C($2_2^+$) at
$E$ = 2.6 MeV, as illustrated analogously by the potential energy 
landscape of Hill’s three-body problem in Fig.\ 3 of Wheeler \cite{Whe63}. 
Subsequently, the $\alpha$ particle may escape via quantum tunneling 
through the adjacent barriers, which is characterized by overall low penetration
probability, yielding a small $\Gamma_\alpha$ width for the $2_2^+$
resonance.

The $\alpha$-$^8$Be barrier height
depends sensitively on the orientations of $^{8}$Be target nucleus 
\cite{Won73}. In this context, Fig. 2(b) of Biashya {\it et al.}
\cite{Bai21} shows that the energy level of the $0_2^+$ Hoyle resonance is substantially below the potential
barrier. Additionally, their analysis shows that the $2_2^+$ resonance is at
least 1.5 MeV below the $L$ = 2 potential barrier top, which is
positioned at approximately $E$ = 4.3 MeV in their described ``equal
energy" or equilateral triangle (DDE) 3$\alpha$ configuration. Conversely, 
in their ``linear chain" (DDL) 3$\alpha$ configuration, 
Fig. 2(b) indicates that the $2_2^+$ resonance is near the
top of the barrier, though still at least 0.3 MeV below the barrier 
height of $E$ $\approx$ 3.1 MeV.  

From the  viewpoint of the potential landscape, the energy level of
the $2^+$ resonance at $E_x~\approx 10$ MeV (or $E_{\rm c.m.}~\approx$ 2.6 MeV) in question 
 relative to the barrier height
 is either slightly below or at the 
top of the potential barrier  for different escape paths,  and 
the resonance 
would likely have a $\Gamma_\alpha$ below the MeV range.  
We can understand why  the coupled channel potential scattering theory gives a narrow $\alpha$-width of 27 eV.    
It is therefore difficult to understand the physical origin of  the large MeV $\alpha$-width at $E_x~\approx$ 10 MeV  as reported in Ref. \cite{Zim13}.

\section{Evaluation of the cross section for $\alpha$+$^8$Be to fuse 
and decay to the $^{12}$C($2_1^+)$}

In $\alpha$+$^8$Be$_{\rm gs}$ collisions at  $E = E_{\rm c.m.} \lesssim 3.03$ MeV, below the energy 
for the excitation of the $^8$Be($2^+$) state, we can enumerate the various possible reaction channels. 
The first is the elastic channel in which both $\alpha$ and $^8$Be remain in their ground states. 
The second involves the reaction channels associated with the $L$ partial-wave, where $\alpha$ and $^8$Be 
nuclei form a $^{12}$C*($L^\pi$) compound nucleus in a resonance of a continuum state with 
angular momentum $L$, resulting in a total reaction cross section $\sigma_R(E, L)$. 

The reaction cross section can be further divided into partial reaction cross sections 
for the compound $^{12}$C system with angular momentum $L$ to decay electromagnetically.
This includes either cascading to the ground $^{12}{\rm C}(0_1^+)$ state with a $\gamma$-emission 
width $\Gamma_\gamma({}^{12}{\rm C}^*\buildrel \gamma \over\to {}^{12}{\rm C}(0_1^+))$ 
or to the $^{12}{\rm C}(2_1^+)$ state with a width $\Gamma_\gamma({}^{12}{\rm C}^*\buildrel \gamma \over\to {}^{12}{\rm C}(2_1^+))$.  

We can estimate the radiative fusion (RF) cross section, $\sigma_{RF}(E,L)$, for an $\alpha$ 
particle in the $L$ partial-wave state fusing with the $^8$Be$_{\rm gs}$ nucleus to form $^{12}$C
by equating it to the reaction cross section $\sigma_R(E)$, multiplied by the 
radiative fusion branching fraction (or simply radiative fusion fraction) $F(E,L)$ for the 
compound $^{12}$C nucleus decays radiatively to the $^{12}$C$(2_1^+)$ state or $^{12}$C$_{\rm gs}$, that is
\begin{eqnarray}
\sigma_{RF}(E,L)= \sigma_R(E,L) F(E,L ).
\label{fu}
\end{eqnarray}

In heavy-ion collisions the Bethe model of strong absorption in-going wave 
boundary approximation \cite{Bet40} assumes $F(E,L)$ = 1, leading to the commonly accepted 
approximation of $\sigma_F=\sigma_{RF}$ \cite{Won73}. This arises because, in the context of heavy-ion collisions, 
the masses of the colliding nuclei are so large and the probability of tunneling of 
large-mass nuclei from the fused system back to the dissociated system below the barrier 
is so low compared to the probability for radiative decay to lower energy states, which 
are numerous in number. Consequently, the tunneling probability of a fused two-heavy-ion system 
is small in comparison to the probability of radiative decay, resulting in the radiative fusion fraction $F(E,L)$ 
being close to 1. Thus, 
the 
radiative fusion cross section, which can be called simply the fusion cross section $\sigma_F$,
is approximately equal to the 
reaction cross section,  $\sigma_{RF}\equiv \sigma_{F} ~\approx ~\sigma_R$.  

However, this relationship of $F(E,L)~\approx$ 1 may not hold for light nuclei, 
which have few final excited states onto which the compound nucleus excited 
state can de-excite by $\gamma$-emission. The $F(E, L)$ 
could be substantially less than 1.  It is necessary to evaluate the 
above $\Gamma_\gamma$ and $\Gamma_\alpha$ as a function of energy for the resonances near 
the Hoyle resonance. In other words, the in-going wave approximation is reasonable 
for heavy-ion collisions with large heavy-ion mass because tunneling for a 
massive projectile through a barrier is difficult, but it may not be applicable 
for a light-ion projectile such as an $\alpha$-particle. 

\begin{figure}[h]
\centering
\includegraphics[scale=0.4]{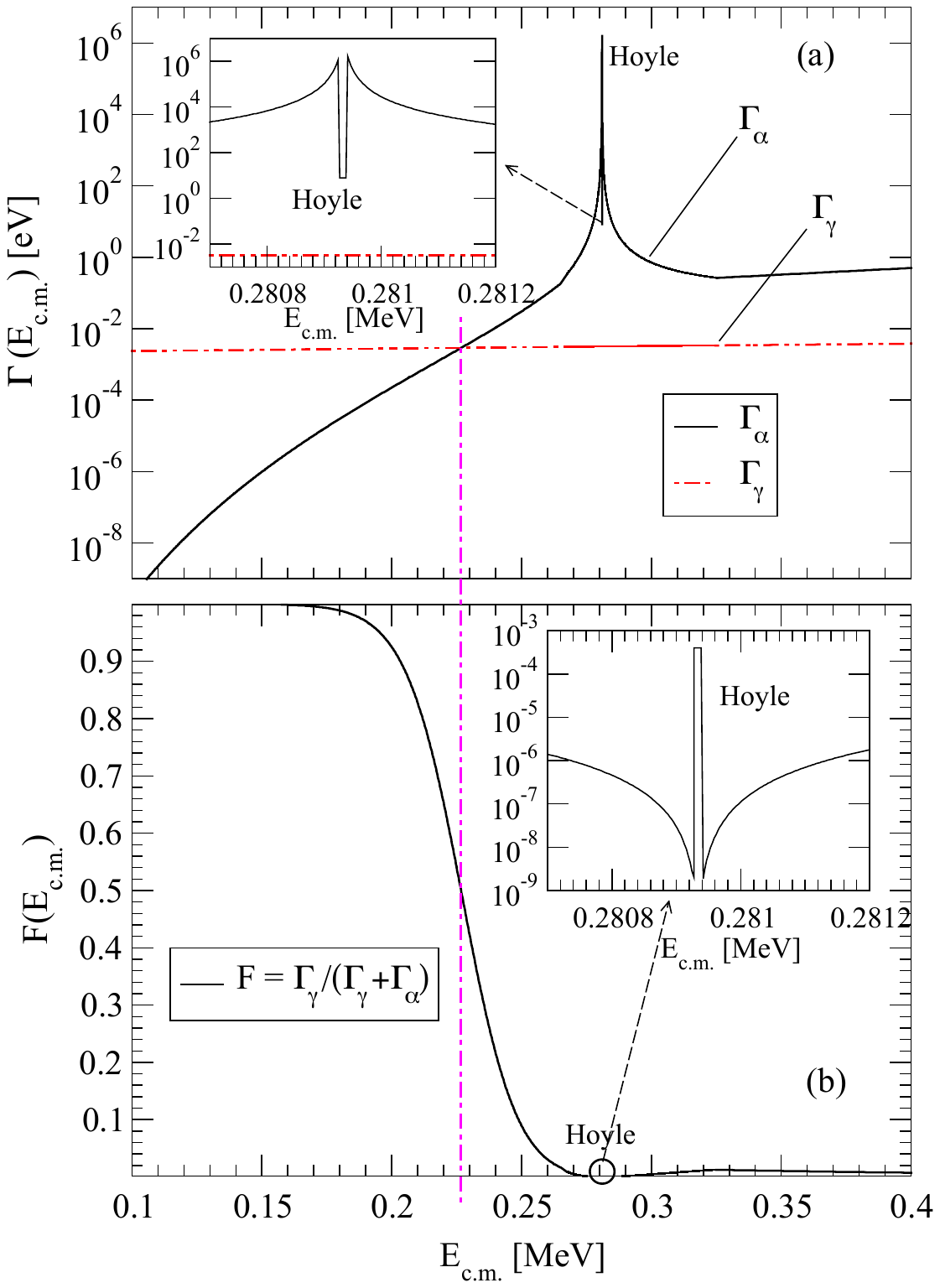}
\vspace{-1cm}
\caption{
(a) Energy dependence of $\Gamma_{\alpha}(E_{\rm c.m.})$ and $\Gamma_{\gamma}(E_{\rm c.m.})$ of the $^{12}$C compound nucleus state formed in the $\alpha$+$^8$Be 
collisions. (b) Energy dependence of radiative fusion fraction $F(E_{\rm c.m.})$  for $s$-wave.}
\label{fig6}
\end{figure}

We shall henceforth focus on the lowest energy region where $E < 1$ MeV 
for which $L = 0$, and we omit the label $L = 0$. In this case, as pointed out in 
Ref.\cite{Fre14},  $\Gamma_\gamma(  ^{12}{\rm C}(0_2^+) \buildrel \gamma \over\to{} ^{12}{\rm C}(2_1^+) \gg \Gamma_\gamma( ^{12}{\rm C}(0_2^+) \buildrel \gamma \over \to
{}^{12}{\rm C}(0_1^+) $, so it is reasonable to ignore the radiative transition of \break $^{12}{\rm C}(0^{+*},E) 
\buildrel \gamma \over 
\to {} ^{12}{\rm C}(0_1^+)$ in comparison with the $E2$ transition of $^{12}{\rm C}(0^{+*},E )
\buildrel \gamma \over 
\to  ^{12}\!\!{\rm C}(2_1^+)$. Thus, the total radiative width of the $^{12}$C nucleus in the continuum, $\Gamma_{\gamma}(E)$, is just given approximately by $\Gamma_\gamma( ^{12}{\rm C}(0^{+*},E) 
 \buildrel \gamma \over 
 \to {}^{12}{\rm C}(2_1^+))$. 
 
In addition to the radiative decay, the compound  nucleus  $^{12}$C$^*(E)$ can also decay via the emission 
of an $\alpha$ to return to the state of $\alpha$+$^{8}$Be$_{\rm gs}$ nucleus with a width
$\Gamma_\alpha(({}^{12}{\rm C}^*,E)\buildrel\alpha\over \to {}^8{\rm Be} + \alpha,E)$.
The radiative fusion fraction $F(E)$ in this lowest energy region is therefore
\begin{eqnarray}
\!\!F(E )&&=
 F(^{12}{\rm C}(0^{+*},E )\buildrel \gamma \over\to {}^{12}{\rm C}(2_1^+)) 
\nonumber\\
&&\hspace{-0.0cm}=\frac{\Gamma_\gamma({}^{12}{\rm C}(0^{+*},E)\buildrel \gamma \over\to {}^{12}{\rm C}(2_1^+))}
{\Gamma_{\rm total}({}^{12}{\rm C}(0^{+*},E))},
\label{eq5}
\end{eqnarray} 
where
\begin{eqnarray}
\!\!&&\!\!\Gamma_{\rm total}({}^{12}{\rm C}(0^{+*},E))
\nonumber\\
&&
\!\!=\!{\Gamma_{\gamma}({}^{12}{\rm C}(0^{+*},E\!)\!\!\buildrel \gamma \over\to\!\! {}^{12}{\rm C}(2_1^+))\!+\! \Gamma_\alpha({}^{12}{\rm C}(0^{+*},E\!)\!\!\buildrel\alpha\over \to\!\! {}^8{\rm Be} \!+ \!\alpha).}
\nonumber\\
\end{eqnarray}

The $^{12}$C($2_1^+$) state will eventually decay down to its ground state, with a fused stable $^{12}$C nucleus. 
The above quantity gives $F(E)$ and $\sigma_F(E)=F(E)\sigma_R(E) $ is therefore the $s$-wave $^{12}$C 
fusion cross section, in the collision of the $\alpha$-particle with $^8$Be$_{\rm gs}$.

To calculate $\Gamma_\gamma$, we make use of the energy dependence 
of the $\gamma$-ray for $E2$ transition. We surmise that the values of the $E2$ matrix elements 
are approximately the same for the Hoyle resonance and those of nearby resonances close to the 
Hoyle, because they are in the same neighborhood, and the energy difference between them 
is small compared to the energy difference for the $\gamma$ decay 
$E_x(^{12}{\rm C}(0^{+*},E)- E_x({}^{12}{\rm C}(2_1^+))$, where $E_x (^{12}{\rm C}(0^{+*},E))
= E+7.367$ MeV. Therefore \cite{Bla52},  
\begin{eqnarray}
&&\Gamma_\gamma (E) 
\nonumber\\
&&= \frac{4[E_x({}^{12}{\rm C}(0^{+*},E))- E_x(^{12}{\rm C}(2_1^+))]^5\pi}{15}B(E2; 2_1^+ \to 0_2^+),
\nonumber\\
\label{gam}
\end{eqnarray}
where $E_x(^{12}{\rm C}(2_1^+))$ = 4.4398 MeV 
and $B(E2; 2_1^+ \to 0_2^+)$ = 2.7 $e^2$ fm$^4$ is the experimental $E2$ strength 
from \cite{Sel90}.  According to Table I of \cite{suno15},
the theoretical $B(E2; 2_1^+ \to 0_2^+)$ values can vary from approximately 1 to 10 $e^2$ fm$^4$, depending on 
the theoretical methodologies and approximations made within the calculations.

We further assume that the penetrability $P_{\alpha}(E)$, for $\alpha$-particle returning to 
the entrance channel is the same as for the $\alpha$-particle to tunnel into the potential 
barrier in the potential scattering description, therefore we can express our $\Gamma_\alpha$ for $\alpha$ 
escaping back to the exit channel as
\begin{eqnarray}
\Gamma_\alpha (E)= \hbar \nu (E) P_{\alpha}(E),
\label{eq7}
\end{eqnarray}
where $\nu$ is the frequency of assault on the potential wall and $P_{\alpha}(E)$
is the penetrability which can be extracted from the reaction cross sections, 
whose energy dependence is shown in Fig.\ \ref{fig3}(b). 
That is, for our case when we limit our attention to $L = 0$ wave, $P_\alpha(E)$ 
is $P_{\rm res}(E, L=0)$ of Eq.\ (\ref{BWF}) extracted from the results of potential 
scattering theory in the coupled-channel calculations
at the resonance shown in Fig.\ \ref{fig3}(b), and  $P_\alpha(E, L=0)$ 
is $P_{\rm contin}(E)$ of Eq.\ (\ref{contin}) for energies of the off-resonance 
in the continuum region extracted from the $\sigma_R$ of Fig.\ \ref{fig3}(b).

The $s$-wave potential curve has a peculiar shape as a function of $r$ requires us 
to obtain the  frequency of assault
$\nu$
through the period $T(E)$, in units of fm/$c$, given by
\begin{eqnarray}
T(E) =\frac{1}{\nu (E)}= \int _{R_T(E)}^0 \frac{2dr}{\sqrt {2(E - V(r))/\mu}},
\end{eqnarray}
where $R_T$, in units of fm, is the turning point satisfying $E=V(R_T)$ with the reduced mass $\mu$ in units of MeV. 

\begin{figure}[h]
\centering
\includegraphics[scale=0.34]{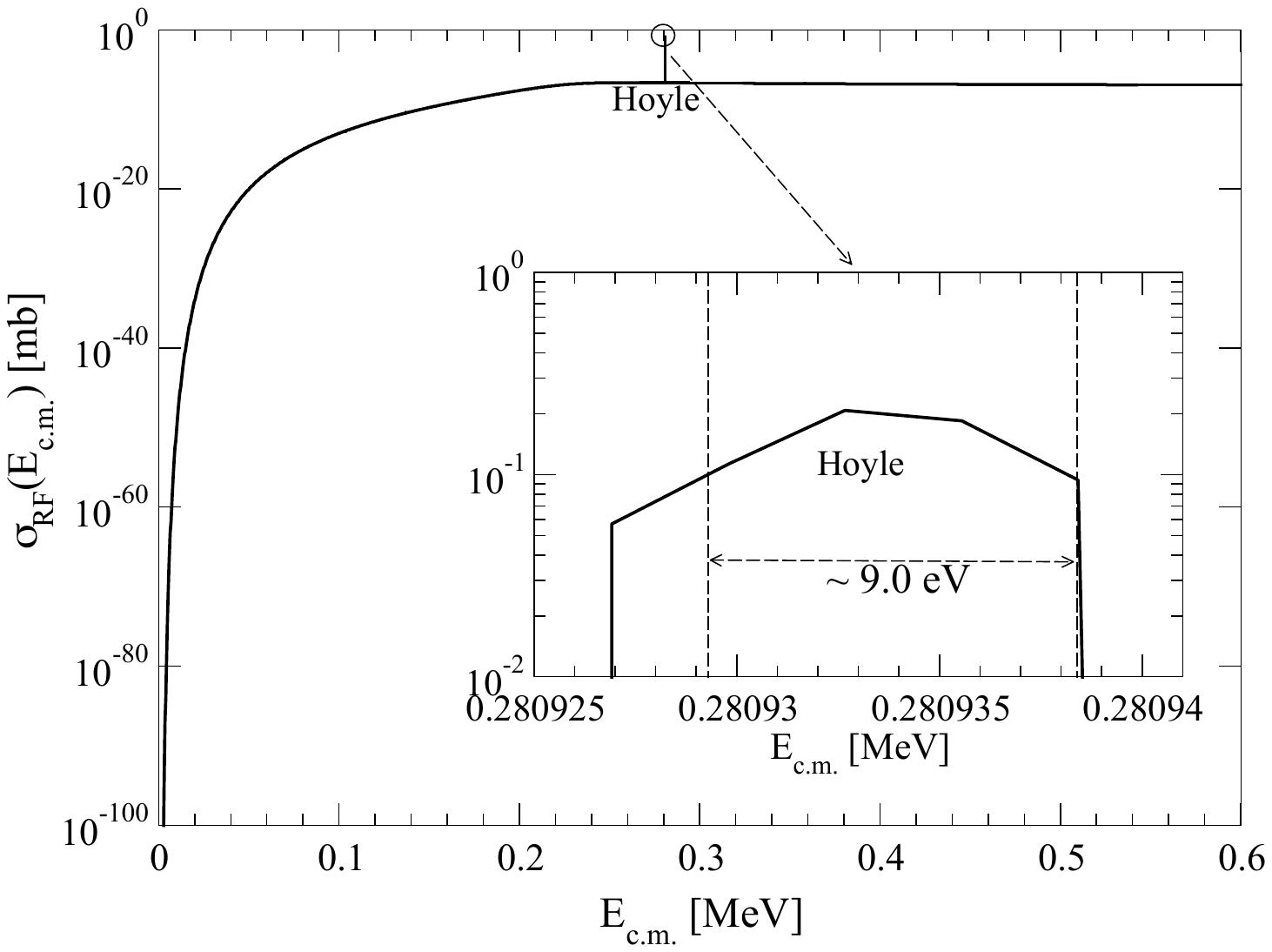}
\vspace{-0.75cm}
\caption{
Semi-logarithmic plot of the radiative fusion cross section for $\alpha + {}^8{\rm Be} \to {}^{12}{\rm C}(2_1^+)$.
The red circles show that the cross section near the threshold region can be described by Eq.\ (\ref{thresholdlaw}).}
\label{fig7}
\end{figure}

Figure \ref{fig7} shows the radiative fusion cross section of Eq.\ (\ref{fu}) 
for $\alpha + {}^8{\rm Be} \buildrel \gamma \over\to {}^{12}{\rm C}(2_1^+)$ as a function of energy. 
The plot clearly shows the Hoyle resonance as a sharp, well-defined spike on the broad Gamow-like background. A closer view of the Hoyle region, 
provided in the inset, indicates $\Gamma_\alpha$ $\approx$ 9.0 eV, in agreement with 
the previously calculated width of 8.0 eV 
and consistent with 
the experimental value 
of 8.5 eV within uncertainty. In the near-threshold region where $E_{\rm c.m.}$ $\lesssim$ 0.12 MeV,  
the $\sigma_{RF}$ cross section appears to follow effectively the relation
\begin{equation}
 \ln(\sigma_{RF}(E_{\rm c.m.})/{\rm mb}) = A \ln\{B \ (E_{\rm c.m.}/{\rm MeV})^D+ C\}, 
 \label{thresholdlaw}
 \end{equation}
where the coefficients $A = 55.45$, $B = 2.646$, $C$ = $-$1.459 and $D$ = 0.112. 

The energy-dependent product  ($\sigma${\it v}) for the electric quadrupole ($E2$) transition to the 
${}^{12}{\rm C}(2_1^+)$ state was calculated by Ogata {\it et al.} \cite{Oga09} using the continuum-discretized 
coupled-channel (CDCC) method to solve the three-body Schr\"{o}dinger equation, accounting 
for both resonant and nonresonant contributions to the triple-$\alpha$ reaction at low temperatures. 
They called this  ($\sigma${\it v}) quantity  ``the reaction probability''. Suno {\it et al.} \cite{suno15} 
subsequently applied the adiabatic hyperspherical complex absorbing potential (HCAP) method to obtain 
the energy-dependent photodisintegration cross section $\sigma_{\gamma}$ 
for ${}^{12}{\rm C}(2_1^+)+\gamma$ $\to$ $\alpha+\alpha+\alpha$, finding 
good agreement with the results of Nguyen $et~al.$ \cite{Tho12} and Ishikawa \cite{Ish13}.

\begin{figure}[h]
\centering
\includegraphics[scale=0.35]{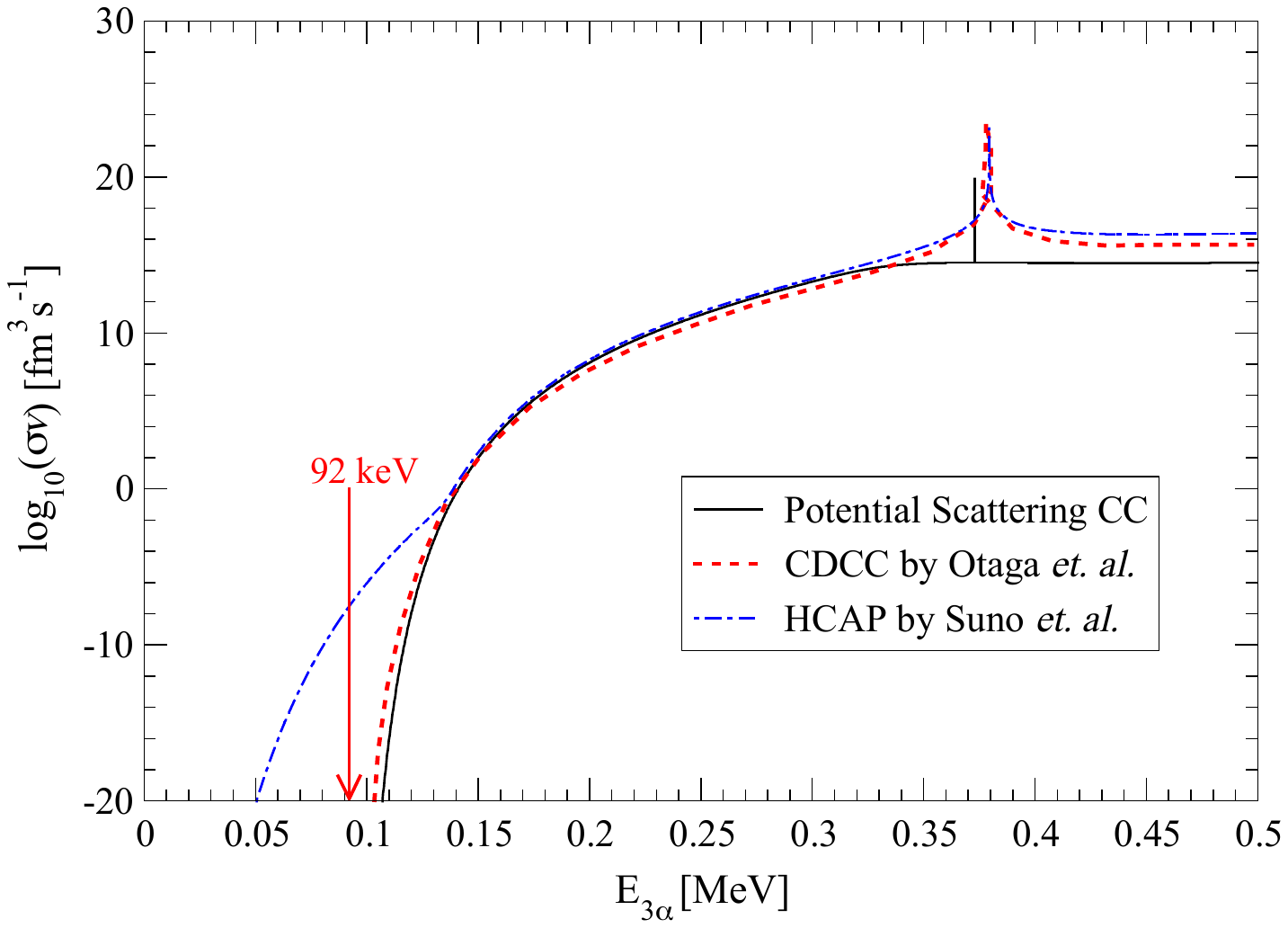}
\vspace{-0.6cm}
\caption{ 
Comparison of the reaction probability  ($\sigma\!_{_{RF}}${$v_{cm}$}) 
for the radiative fusion process $\alpha + {}^8{\rm Be} \to {}^{12}{\rm C}(2_1^+)+\gamma$ with the reaction probability  ($\sigma${\it v})
 for the triple-$\alpha$ 
 radiative capture process,
 $\alpha+\alpha+\alpha$ $\to$ ${}^{12}{\rm C}(2_1^+)+\gamma$.  Here, the solid curve gives 
  ($\sigma\!_{_{RF}}${\it v}) for the $\alpha$-$^8$Be radiative fusion process $\alpha + {}^8{\rm Be} \to {}^{12}{\rm C}(2_1^+)+\gamma$,
obtained from potential scattering theory coupled‑channel calculations,
with $E_{3\alpha}=E_{\rm c.m.}(\alpha^8{\rm Be})+2M(\alpha)c^2-M(^8{\rm Be})c^2=E_{\rm c.m.}(\alpha^8{\rm Be})
+92.08$ keV.
 The threshold energy $E_{3\alpha} =$ 92.08 keV for the onset of the radiative fusion of $\alpha+{}^8$Be  to $^{12}$C is indicated by an arrow.
 Results of CDCC in the dotted curve and HCAP in the dashed-dotted curve  are taken from Ref.\cite{Oga09} and Ref.\cite{suno15}, respectively.}
\label{vsig}
\end{figure}
 
It is instructive to compare our  results  with those  from previous 
triple-$\alpha$ calculations.  In such comparisons, 
it should however be kept in mind that  our model, 
 which follows Salpeter and Hoyle fusion mechanism as described in the Introduction, 
  is a two-body scattering model between an $\alpha$ and $^8$Be$(0^+)$.
   Its domain of applicability is the region of $E_{\rm c.m.}>0$ for a system  with preformed ${}^{8}$Be, such as a medium of  $\alpha$ particles  in thermal equilibrium with $^8$Be$(0^+)$ nuclei at temperature $T$.
On the other hand, the triple-$\alpha$  calculations employ  a three-$\alpha$-particle   description, which includes, besides 
   the important channel of an $\alpha$ colliding with an ($\alpha$-$\alpha$) cluster that may be considered as an incipient or preformed $^8$Be,  also  the additional three-particle phase space, and three-particle effects of alpha particle exchange symmetry \cite{Oga09,suno15,Tho12,Ish13}, as well as possible three-body forces \cite{suno15, Tho12,Ish13}.
  The additional three-particle effects are expected to be important at low energies in the region of small $E_{3\alpha}$ and not in  the region of high $E_{3\alpha}$.  Consequently,  the region of high $E_{3\alpha}$ energies will be likely dominated by the fusion of an $\alpha$ particle with an incipient or preformed $^8$Be nucleus. We would expect that our results from the two-body treatment in the present framework should be close to the results of the triple-$\alpha$ calculations from the three-body treatment at  high  $E_\alpha$  but will not be applicable in situations where the three-body effects are expected to be important, as in the low $E_{3\alpha}  $ region.

To make such a comparison, we show in Fig. \ref{vsig} our reaction probability, ($\sigma\!_{_{RF}}${$v_{cm}$}),  
for the radiative fusion process $\alpha + {}^8{\rm Be} \to {}^{12}{\rm C}(2_1^+)+\gamma$, together with the reaction probability ($\sigma${\it v})
for the triple-$\alpha$ radiative capture process $\alpha+\alpha+\alpha$ $\to$ ${}^{12}{\rm C}(2_1^+)+\gamma$, as inferred from the numerical CDCC results of Ogata $et~al$. \cite{Oga09}, as a function of the collision energy $E_{3\alpha}$, where $E_{3\alpha}=E_{\rm c.m.}$ + 92.08 keV, and $v_{cm}$ = $\sqrt{E_{\rm c.m.}/2\mu_{\alpha+{}^{8}{\rm Be}}}$. 
To place the two calculated quantities on an equal footing for comparison, we examine the 
formulations of reaction probability in photodisintegration cross section 
presented in \cite{suno15} and \cite{Oga09}, apply the reciprocity theorem, multiply
the velocity $v_{3\alpha}$ derived from $E_{3\alpha}$, and introduce 
a multiplicative factor $2^6\pi^4$ to the photodisintegration cross section formula
arising from the normalization factor for the three-$\alpha$ scattering wave function in free
space (see details in Ref.\cite{Oga09} and Ref.\cite{suno15}).
As shown in Fig.\ \ref{vsig}, the results of our ($\sigma\!_{_{RF}}${$v_{cm}$})  and those of  ($\sigma${\it v}) from the  CDCC results of Ogata $et~al.$ \cite{Oga09}, using pairwise two-body interactions, are in general approximate agreement with each other for a large range of energies.  Such an agreement indicates that for energies $E_{3\alpha} \ge 92.08$ keV, the $\alpha ^{8}$Be fusion process dominates the reaction probability of three $\alpha$ particles fusing into a $^{12}$C nucleus. In Fig.\ \ref{vsig} we also compare our reaction probability with that of Suno $et~al.$ \cite{suno15}, whose calculation includes an additional attractive three‑body interaction among the $\alpha$ particles.  
As expected, this additional three $\alpha$ particles interaction enhances the fusion reaction probability 
at low $E_{3\alpha}$, relative to those of triple-$\alpha$ CDCC results  which employ only pairwise two‑body interactions, and also relative to our $\alpha$+$^8$Be radiative fusion results.

\begin{figure}[h]
\centering
\includegraphics[scale=0.35]{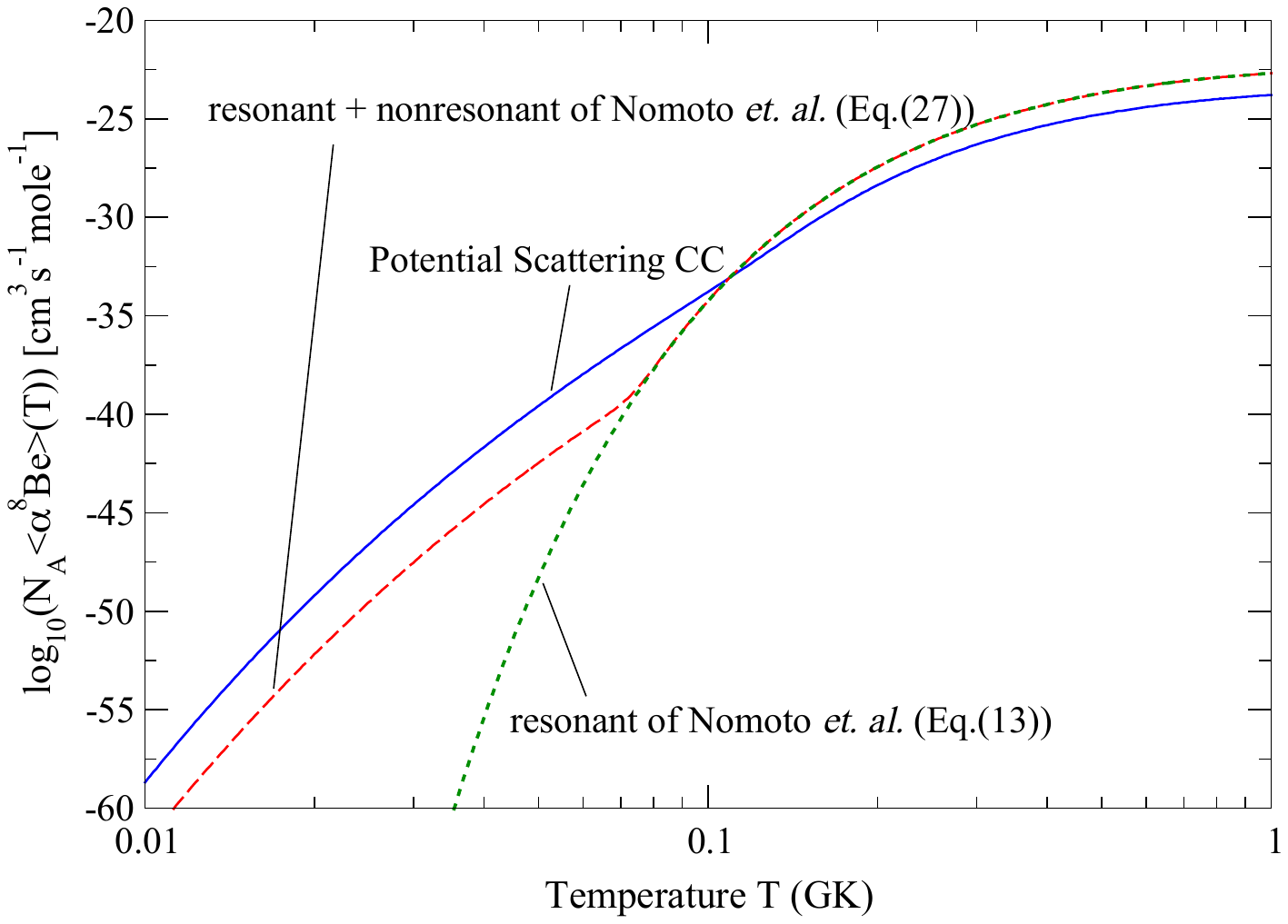}
\vspace{-0.6cm}
\caption{
The plot compares the Maxwellian-averaged $\alpha$+$^8$Be radiative fusion (capture) rate, $N_A\langle \alpha ^8\text{Be} \rangle$ [cm$^3$/s/mole],  
as a function of temperature $T$ [GK] from the potential scattering theory coupled‑channel calculations as the 
solid curve, with those from Eqs.\ (13) and (27) of Nomoto {\it et al.} \cite{Nom85} as the dashed and the dotted curves, respectively.
The solid curve and dashed curves include both resonant and nonresonant contributions, while the dotted curve of includes only the 
resonant contribution.}
\label{rates}
\end{figure}

As another check of our theoretical radiative fusion cross section, we consider
 a medium of $\alpha$ particles at a temperature $T$ in thermal 
 equilibrium with $^8$Be nuclei. The probability distribution of the $^8$Be nuclei is 
 given by a Maxwell-Boltzmann distribution characterized by the temperature $T$. 
 The radiative fusion (capture) rate $N_A \langle \alpha ^8{\rm Be}\rangle$, expressed in units of cm$^3$/s/mole for an $\alpha$ colliding 
 with $^8$Be to form $^{12}$C, can be obtained by integrating the radiative-fusion probability element over the energies, 
\begin{eqnarray}
N_A \langle \alpha ^8{\rm Be}\rangle 
=N_A \sqrt{\frac{8}{\pi \mu (k_BT)^3}}\!\!\int_{E_i}^{E_f} \hspace{-0.8em} \nonumber
\sigma_{_{RF}}(E) E e^{(-E/k_BT)} dE, \\
\label{ratesT}
\end{eqnarray}
where $N_A$ is the Avogadro's number, $\mu$ is the reduced mass of the colliding particles, $k_B$ is the Boltzmann constant 
and $E_{(i,f)}$ denote the relevant energy limits. In Fig. \ref{rates} we compare the radiative fusion rate  $N_A \langle \alpha^8\text{Be}  \rangle$ for the reaction $\alpha +^8\text{Be}\to {}^{12}{\rm C}(2_1^+)+\gamma$ of Eq.\ (\ref{ratesT}),
obtained from the potential‑scattering coupled‑channels (CC) theory (solid curve), with the 
radiative capture rate  $N_A \langle \alpha ^8\text{Be}  \rangle$ 
for the reaction  $\alpha + {}^8{\rm Be} \to {}^{12}{\rm C}(2_1^+)+\gamma$ given by Eq.\ (27) of Nomoto {\it et al.} \cite{Nom85} (dashed curve), which includes both resonant and nonresonant contributions. 
The two calculations agree with each other to within two orders of magnitude.

\begin{figure}[htp]
\centering
\includegraphics[width=1.0\linewidth,height=0.75\linewidth]{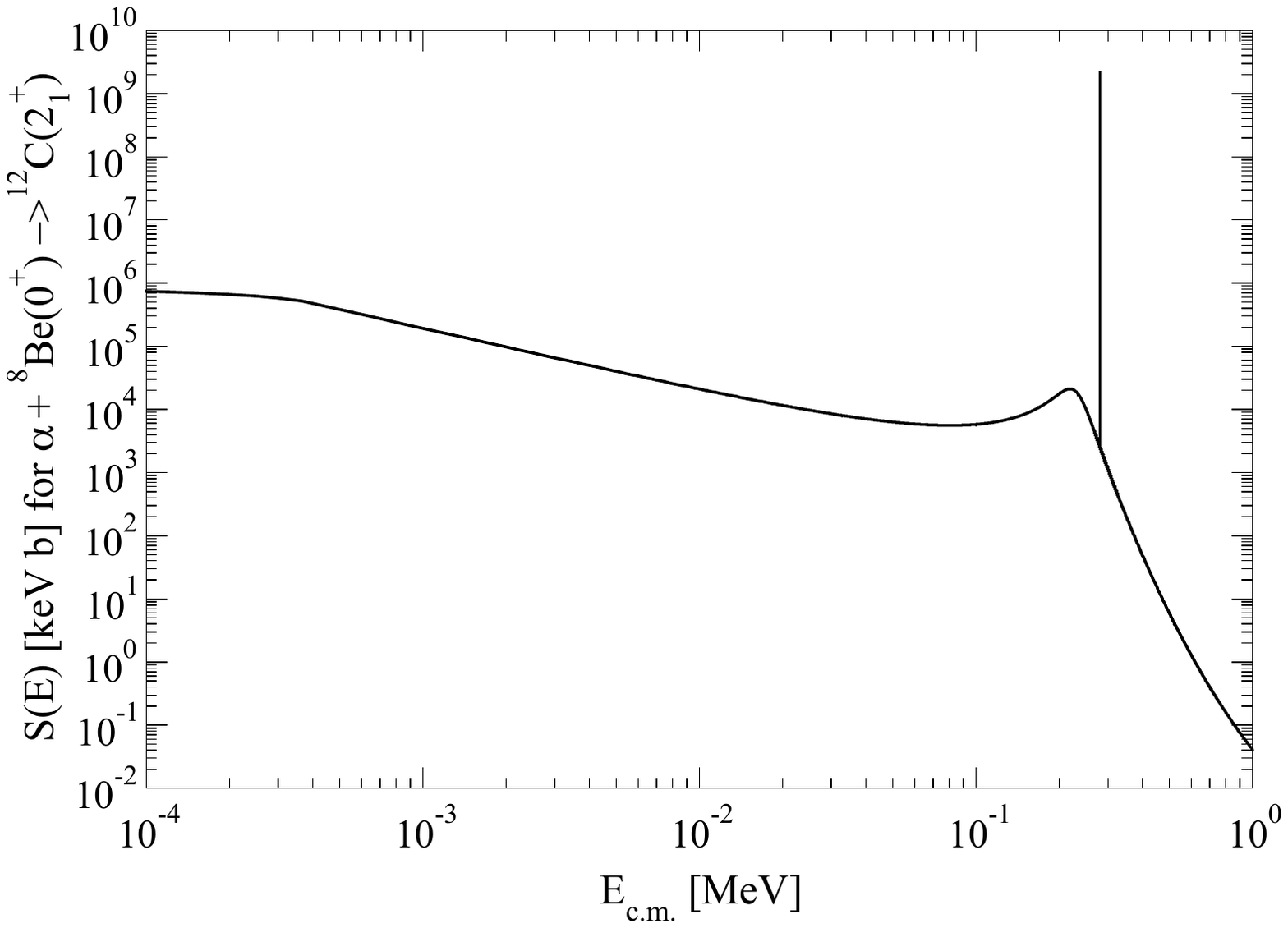}
\vspace{-0.5cm}
\caption{The astrophysical $S(E)$ factor for $\alpha + {}^8{\rm Be} \to {}^{12}{\rm C}(2_1^+)$ 
radiative fusion reaction as a function of energy.}
\label{fig8}
\end{figure}

\begin{figure}[h]
\centering
\includegraphics[scale=0.35]{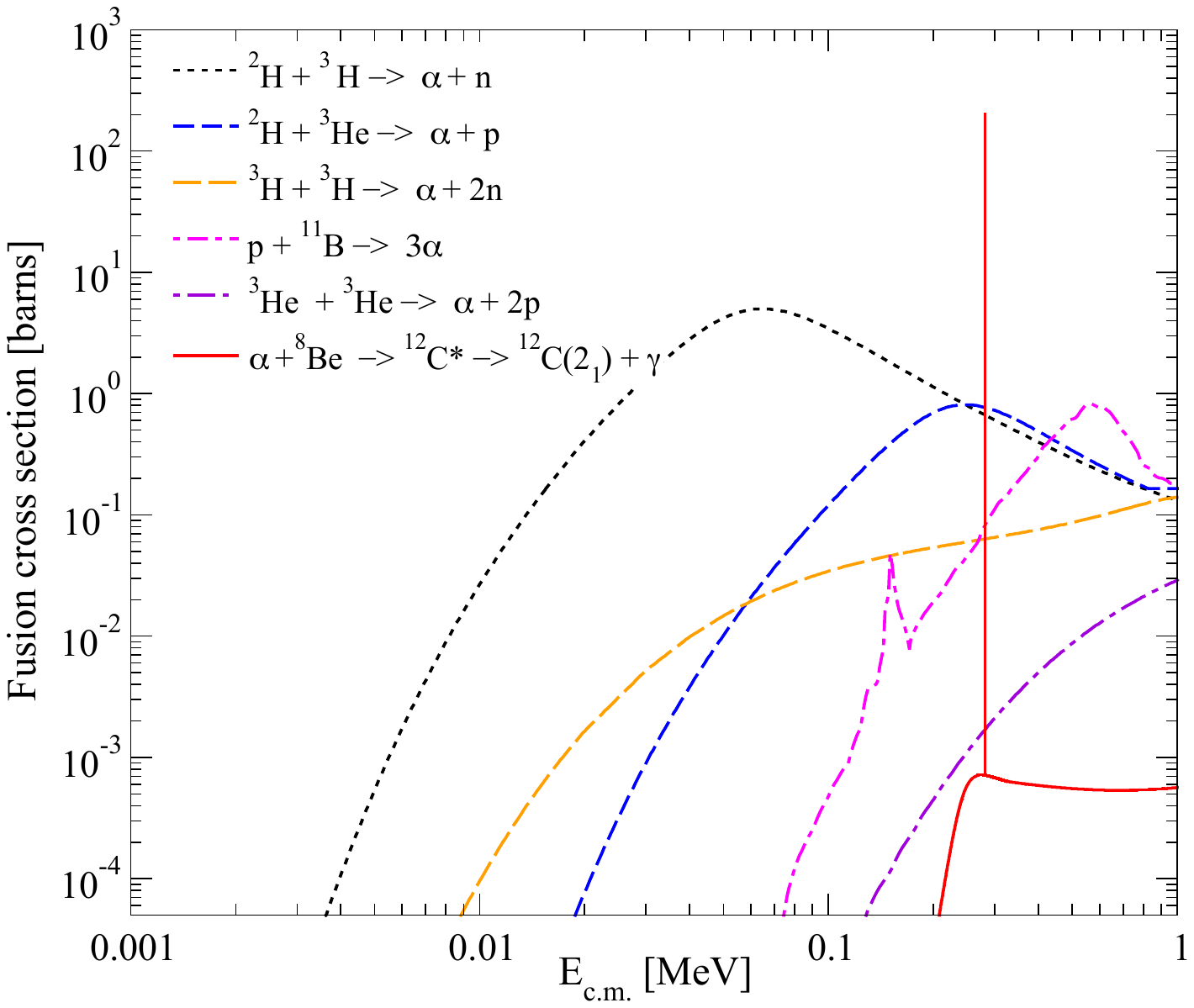}
\vspace{-0.6cm}
\caption{ 
The collection of the fusion cross section data for many $A \le$ 12 systems.
The recommended fusion data are from the IAEA's ENDF \cite{endf}, except 
for the $p +^{11}$B reaction data which come from the JANIS \cite{janis} database.
}
\label{fig9}
\end{figure}

For completeness, we also evaluated 
the corresponding astrophysical $S$-factor, $S(E)$ = $\sigma\!_{_{RF}}E\exp(2\pi\eta)$, as shown in Fig.\ \ref{fig8}.
Furthermore, it is  informative to compare our estimated radiative fusion cross section for $\alpha + {}^8{\rm Be}$ 
with the five well-known fusion reactions in Fig.\ \ref{fig9}. Our radiative fusion cross 
section resembles, in particular, the general behavior of either the D+T or D+$^3$He, or $p$+$^{11}$B reactions, 
and is by far the smallest among the samples considered here, except at the Hoyle resonance.

We should examine how the decay of the $^8$Be($2^+$) nucleus will affect the reaction cross section 
at energies below 4 MeV. During the collisions, the $2^+$  resonance can be excited virtually from 
the $^8$Be$_{\rm gs}$ and then return to the ground state, leading to a polarization potential through the 
second-order perturbation theory \cite{Hus84}. This polarization potential is accounted for 
in the coupled-channel approach. Virtual excitation does not involve the disintegration of the $^8$Be state 
into two $\alpha$ particles, whereas real excitation leads to the population of the unstable 2$^+$ state, 
which subsequently disintegrate into two $\alpha$ particles by 1.513 MeV \cite{Til04}. 
However, the $^8$Be($2^+$) channel remains energetically close and insignificant for 
the projectile energy in question with respect to 3.03-MeV, the $^8$Be($2^+$) resonance energy. 
Thus, using the coupled-channel method calculated inelastic cross section $\sigma(E) = \sigma^{\rm tot}_R(E)+\sigma_{\rm ex}(E)$ and $^8$Be$( 2^+)$ 
excitation cross section $\sigma_{\rm ex}(E)$ with $\Gamma(^8{\rm Be}(2^+\hspace*{-0.1cm}\rightarrow\hspace*{-0.1cm}0_1^+), \gamma)$ = 8.3 meV, $\Gamma(^8{\rm Be}(2^+)\hspace*{-0.1cm}\rightarrow\hspace*{-0.1cm}2\alpha)$ = 1.513 MeV and
${\Gamma(^8{\rm Be}(2^+)\hspace*{-0.05cm}\rightarrow \hspace*{-0.05cm}2\alpha)}/{\Gamma_ {\rm total}(^8{\rm Be}(2^+))}$
$\approx$ 1.0, where\break  $\Gamma_ {\rm total}(^8{\rm Be}(2^+))$ is the total width for $\alpha$ and $\gamma$ decays, 
we evaluated the $^8$Be($2^+$) breakup fraction,
\begin{eqnarray}
F_{2^+{\rm breakup}}
= \left(\frac{\sigma_{\rm ex} (E)}{\sigma(E)}\right)
\frac{\Gamma(^8{\rm Be}(2^+)\hspace{-0.05cm}\rightarrow\hspace{-0.05cm}2\alpha)}{\Gamma_ {\rm total}(^8{\rm Be}(2^+))},
\end{eqnarray}
and found them to be small for collision energies up to 4.0 MeV. For example, 
the fraction $F_{2^+{\rm breakup}}$ is on the order of 10$^{-2}$ at $E_{\rm c.m.}$ = 4 MeV.

\section{Conclusions and Discussions}
We have used the potential scattering theory 
in the coupled-channel framework to study the Hoyle-Salpeter process of the collision between 
$\alpha$ and  $^8$Be ground state nucleus in the Hoyle resonance and associated resonances
region. The $^{12}$C resonance energies  and their corresponding $\alpha$-widths have 
been used to constrain the nuclear potential between the colliding nuclei.

The constraint of the $^{12}$C  experimental  data  in the Hoyle 
resonance
region reveals that the nuclear potential between $\alpha$ and
$^8$Be needs to include a parity-dependent surface potential component such that it is
more attractive for even-$L$ positive-parity partial waves than for
odd-$L$ negative-parity partial waves. As a consequence, the radial
dependence of the total potentials for $^{12}$C $\{0^+ ,2^+ ,4^+ \}$
states exhibit a double-hump behavior, possessing two local energy
minima and two $^{12}$C $\{0^+ ,2^+ ,4^+ \}$ states with wave
functions centered around two different energy minima, around the
Hoyle state energy region.  The lower members of the  doublet of each of the 
$\{0^+,2^+,4^+\}$ states need to pass through a greater potential
barrier and hence possess narrower $\alpha$-widths compared to their
corresponding higher-energy counterpart doublets.  The lower $0_2^+$ 
doublet of $\{0^+,2^+,4^+\}$ set may be  the  well known  $0_2^+$ Hoyle state. 
However, the lower theoretically predicted $2_2^+$ state at $E_x \approx 10.1$ and
the predicted $4_1^+$ state at $E_x\approx 9.6$ MeV are as yet unobserved
or unidentified.  They need to be searched and uncovered
experimentally to test the double-hump potential description.

The theoretical approach reproduced 
many  resonance energies and widths, including the Hoyle state, 
in approximate agreement with other theoretical calculations and experimental data. 
It complement similar studies on the states of the compound $^{12}$C
system using nuclear structure theories.  While we focus in this
manuscript on the physics of $\alpha+^8$Be, the method 
can be similarly applied to study the states of many
other simple light nuclear systems with few degrees of
freedom.  For example, we can use the method for the collision of
$\alpha^{n_1}$ on $\alpha^{n_2}$ to examine a subset of the $\alpha^{n_2+n_2}$
states that have a large overlap with the colliding nuclei.  By
including the spin algebra, we can also use the present method to
study the collisions of $p$+$^3$H, $p+ ^7$Li, and $p+^{11}$B in
connection with the states of the $^4$He, $^8$Be, and $^{12}$C,
respectively, both at resonances and off-resonances.  The latter
reactions are associated with the production of anomalous bosons with
masses in the tens of MeV \cite{Kra16}, and may be connected with
particles beyond the standard model \cite{Fen16} or QED mesons
\cite{Won10,Won20,Won24}, and may therefore be of considerable
interest.

There is however a limitation in the present method.  The scattering
theory approach  relies on the elastic
and inelastic scattering of a projectile on a target
nucleus.  The operators that connect the initial and final states  have
natural parity and angular momenta.  Therefore only natural parity
angular momenta states can be excited by such operators  and the
scattering theory through the coupled-channel approach can only
explore natural parity states of the combined system; unnatural
parity states of the compound system are beyond the scope of the
present approach. 

From our numerical calculations, we notice the interesting effects of the
interference between the scattering wave propagation and the barrier penetration.  That is,
the barrier penetration does not affect the Breit-Wigner shape of the
resonance in its local energy region.  However, the scattering wave propagation in barrier
penetrability needs to satisfy the boundary condition in the interior of the scattering potential and 
covers an energy regions with an energy  scale large
compared to the width of the Breit-Wigner resonance. {\color{black} Such an interference  modifies the
Breit-Wigner energy profile  into a Breit-Wigner-Fano energy profile, with an additional large energy scale.  
Therefore, the Fano modification may need to be considered in
the application of the Breit-Wigner distribution over an energy region much greater than 
the width of the resonance. }

For astrophysical applications, we examine the relationship between
the reaction and radiative fusion cross sections by calculating the $\gamma$-
and $\alpha$-decay widths for $\alpha$ particles in the potential
pocket. This allows us to assess the $s$-wave radiative fusion cross section and
evaluate the astrophysical $S(E_{\rm c.m.})$-factor for $\alpha + ^8$Be
fusion into $^{12}$C at energies below 1.0 MeV. {\color{black} We have compared our 
coupled‑channel calculations with other previous coupled‑channel studies, 
as well as with radiative fusion rates reported by other authors. 
The comparisons also show reasonable agreement.}

As the study presented here represents one of the first studies on the
fusion of the $\alpha$ particle on the $^8$Be nucleus that is crucial in
nucleosynthesis, we have not yet fine-tuned the various potential
parameters to get a better agreement with the experimental data.  The
agreement with data can be improved by adjusting the potential
parameters. There are also several other improvements to the
theoretical model that can be made for future work.  For example, a
$\beta_4$ deformation parameter can be introduced to better represent
the 2$\alpha$ structure of the $^8$Be nucleus.  Furthermore, in our
comparison with the experimental data of the $^{12}$C($2^+$) state at
$E_x\approx 10$ MeV, we note that the situation is rather complicated.
Much work may need to be carried out to clarify the complexity of the situation.

\vspace*{0.4cm}

\section*{Acknowledgment}
We are indebted to Dr. Ian Thompson of the Lawrence Livermore National
Laboratory for providing guidance, participation and assistance on the
FRESCO code during the initial phase of this work. TGL thanks Hiroya Suno for providing the HCAP results.  
CYW's work is
supported, in part, by the Division of Nuclear Physics,
U.S. Department of Energy under Contract DE-AC05-00OR22725.
\vspace{0.0in}


\end{document}